\documentclass{statsoc}
\usepackage{amsmath}

\usepackage{graphicx,psfrag,epsf}

\usepackage{url} 
\usepackage[textsize=tiny]{todonotes}
\usepackage[left=0.85in,right=0.85in,top=1in,bottom=1in]{geometry}
\usepackage{hyperref}
\hypersetup{colorlinks=true}
\usepackage{xr-hyper}
\newcommand{\calD}{\mathcal D}

\newcommand{\calL}{\mathcal L}

\newcommand{\calU}{\mathcal U}
\newcommand{\calH}{\mathcal H}
\newcommand{\calG}{\mathcal G}
\newcommand{\calO}{\mathcal O}
\newcommand{\eps}{\epsilon}
\newcommand{\given}{\;|\;}

\newtheorem{theorem}{Theorem}

\newcommand{\bX}{\mathbf X}
\newcommand{\bV}{\mathbf V}
\newcommand{\bT}{\mathbf T}
\newcommand{\bW}{\mathbf W}
\newcommand{\bU}{\mathbf U}
\newcommand{\bM}{\mathbf M}
\newcommand{\bN}{\mathbf N}
\newcommand{\bB}{\mathbf B}
\newcommand{\ba}{\mathbf a}
\newcommand{\bm}{\mathbf m}
\newcommand{\bj}{\mathbf j}
\newcommand{\bb}{\mathbf b}
\newcommand{\bv}{\mathbf v}
\newcommand{\bP}{\mathbf P}
\newcommand{\bp}{\mathbf p}
\newcommand{\bq}{\mathbf q}
\newcommand{\bs}{\mathbf s}
\newcommand{\bu}{\mathbf u}
\newcommand{\bh}{\mathbf h}
\newcommand{\bx}{\mathbf x}
\newcommand{\by}{\mathbf y}

\newcommand{\bc}{\mathbf c}
\newcommand{\bw}{\mathbf w}
\newcommand{\bkappa}{\boldsymbol \kappa}
\newcommand{\bgamma}{\boldsymbol \gamma}
\newcommand{\bbeta}{\boldsymbol \beta}
\newcommand{\bOmega}{\boldsymbol \Omega}
\newcommand{\bone}{\mathbf 1}

\newcommand{\bI}{\mathbf I}

\newcommand{\mult}{\mbox{Multinomial} }
\newcommand{\iid}{\overset{iid}{\sim}}
\newcommand{\ind}{\overset{ind}{\sim}}
\newcommand{\jall}{j_1,j_2,\ldots,j_K}
\newcommand{\kpow}{^{(k)}}
\newcommand{\Kpow}{^{(K)}}

\newcommand{\opow}{^{(1)}}
\newcommand{\tpow}{^{(2)}}
\newcommand{\wh}{\widehat}

\newcommand{\tg}{\mbox{Tariff}_\calG}
\newcommand{\tch}{\mbox{Tariff}_\calL}
\newcommand{\tgh}{\mbox{Tariff}_{\calG \cup \calL}}
\newcommand{\tc}{\mbox{Tariff}_{BTL}}
\newcommand{\tnc}{\mbox{Tariff}_{NTL}}
\newcommand{\ig}{\mbox{Insilico}_\calG}
\newcommand{\ih}{\mbox{Insilico}_\calL}
\newcommand{\igh}{\mbox{Insilico}_{\calG \cup \calL}}
\newcommand{\ic}{\mbox{Insilico}_{BTL}}
\newcommand{\inc}{\mbox{Insilico}_{NTL}}
\newcommand{\ej}{\mbox{Ensemble}_{J}}
\newcommand{\ei}{\mbox{Ensemble}_{I}}
\usepackage{etoolbox}

\makeatletter
\patchcmd{\@makecaption}
{\parbox}
{\advance\@tempdima-\fontdimen2} % decrease the width!
{}{}
\makeatother  

\usepackage[caption = false]{subfig} 
\usepackage{enumerate}
\usepackage{comment}
\usepackage{natbib, float}

\title[Bayesian population-level transfer learning]{Regularized Bayesian transfer learning for population-level etiological distributions}
\author[Datta {\it et al.}]{Abhirup Datta}
\address{Department of Biostatistics,
Johns Hopkins University,
Baltimore, 
USA.}
\email{abhidatta@jhu.edu}
\author{Jacob Fiksel}
\address{Department of Biostatistics,
	Johns Hopkins University,
	Baltimore, 
	USA.}
\author{Agbessi Amouzou}
\address{Department of International Health,
	Johns Hopkins University,
	Baltimore, 
	USA.}
\author{Scott L. Zeger}
\address{Department of Biostatistics,
	Johns Hopkins University,
	Baltimore, 
	USA.}

\begin{document}
\begin{abstract}
Computer-coded verbal autopsy (CCVA) algorithms predict cause of death from high-dimensional family questionnaire data ({\it verbal autopsies}) of a deceased individual. CCVA algorithms are typically trained on non-local data, then used to generate national and regional estimates of cause-specific mortality fractions. These estimates may be inaccurate if the non-local training data is different from the local population of interest. 
This problem is a special case of {\it transfer learning} which is now commonly deployed for classifying images, videos, texts, and other complex data. Most transfer learning classification approaches are concerned with individual (e.g. a person's) classification within a target {\it domain} (e.g. a particular population) with training performed in data from a {\it source domain}. Social and health scientists such as epidemiologists are often more interested with understanding etiological distributions at the population-level rather than classifying individuals. The sample sizes of their datasets are typically orders of magnitude smaller than those used for image classification and related tasks. 
We present a parsimonious hierarchical Bayesian {\it transfer learning} framework to directly estimate population-level class probabilities in a target domain, using any baseline classifier trained on source domain data, and a relatively smaller labeled target domain dataset. 
To address the small sample size issue, we introduce a novel shrinkage prior for the transfer error rates guaranteeing that, in absence of any labeled target domain data or when the baseline classifier is perfectly accurate, the {\it domain-adapted (calibrated)} estimate of class probabilities coincides with the naive estimates from the baseline classifier, thereby subsuming the default practice as a special case. A novel Gibbs sampler using data augmentation enables fast implementation. 
We then extend our approach to use not one, but an ensemble of baseline classifiers. Theoretical and empirical results demonstrate how the ensemble model favors the most accurate baseline classifier. 
Simulated and real data analyses reveal dramatic improvement in the estimates of class probabilities from our transfer learning approach. 
We also present extensions that allow the class probabilities to vary as functions of covariates, and an EM-algorithm-based MAP estimation as an alternate to MCMC. An R-package implementing this method for verbal autopsy data is available on Github. 
\end{abstract}

\section{Introduction}\label{sec:intro}
{\it Verbal autopsy} -- a survey of the household members of a deceased individual, act as a surrogate for medical autopsy report in many countries. Computer-coded verbal autopsy (CCVA) algorithms are high-dimensional classifiers, typically trained on non-local data, predict cause of death from these high-dimensional family questionnaires which are then used to generate national and regional estimates of cause-specific mortality fractions (CSMF) elsewhere. These estimates may be inaccurate as  the non-local training data is different from the local population of interest. 

This problem is a special case of {\it transfer learning}, a burgeoning area in statistics and machine learning, on classifying images, videos, text, and other complex data \citep{Pan2010,weiss2016survey}. Models or algorithms trained on {\it source} domain data tend to predict inaccurately in a {\it target} domain different from the source domain in terms of marginal and conditional distributions of the {\it features} (covariates) and {\it labels} (responses) \citep{shimodaira2000improving}. 

Various {\it domain adaptation} strategies have been explored for transfer learning for a generic classification problem which adjust for the difference in marginal and the conditional distributions of the features and labels between the two domains. We refer the readers to \cite{weiss2016survey} and \cite{Pan2010} for a comprehensive review of transfer learning for classification problems. We focus on the setting where there is abundant labeled source domain data, abundant unlabeled target domain data, and limited labeled target data. Transfer learning approaches pertaining to this setting include multi-source domain adaptation \citep[CP-MDA,][]{chattopadhyay2012multisource}, neural networks \citep[TCNN,][]{oquab2014learning}, adaptive boosting \citep[TrAdaBoost,][]{daiboosting,yao2010boosting}, feature augmentation method \citep[FAM,][]{daume2009frustratingly}, spectral feature alignment \citep[SFA,][]{pan2010cross} among others. 

All of the aforementioned transfer learning classification approaches are motivated by applications in image, video or document classification, text sentiment identification, and natural language processing where individual classification is the goal. Hence, they usually focus on the individual's (e.g. a person's or an image's) classification within a {\it target domain} (e.g. a particular population) with training performed in data from a different {\it source domain}. 

Social and health scientists such as epidemiologists are often more interested with understanding etiological distributions at the population-level rather than classifying individuals. 
For example, we aim to estimate national and regional estimates of cause-specific fractions of child mortality. Hence, our goal is not individual prediction but rather transfer learning of population-level class probabilities in the target domain. None of the current transfer learning approaches are designed to directly estimate population-level class membership probabilities. 

Additionally, the extant transfer learning approaches rely on large source domain databases of millions of observations for training the richly-parameterized algorithms. The sample sizes of datasets in epidemiology are typically orders of magnitude smaller than those used for image classification and related tasks. Most epidemiological applications use field data from surveys, leading to databases with much smaller sample sizes and yet with high-dimensional feature sets (survey records). For example, in our application, the feature space is high-dimensional ($\sim 200-350$ features), the `abundant' source domain data has around $\sim 2000$ samples, while the local labeled data can have as few as $\sim 20-100$ samples. Clearly, in such cases, the local labeled data is too small to train a classifier on a high-dimensional set of features, as the resulting estimates will be highly variable. A baseline classifier trained on the larger source domain data will tend to produce more stable estimates, but the high precision will come at the cost of sacrificing accuracy if the source and target domains differ substantially.

Our parsimonious solution to this bias-variance trade-off problem is to use the baseline classifier trained on source-domain data to obtain an initial prediction of target-domain class probabilities, but then refine it with the labeled target-domain data. We proffer a hierarchical Bayesian framework that unifies these two steps. 
With $C$ classes and $S$-dimensional features, the advantage of this new approach is that the small labeled data for the target domain is only used to estimate the $C \times C$ matrix of the {\em transfer error} (misclassification) rates instead of trying to estimate $\calO(SC)$ parameters of the classifier directly from the target-domain data. Since $S\gg C$, this approach considerably reduces the dimensionality of the problem. To ensure a stable estimation of the misclassification matrix, we additionally use a regularization prior that shrinks the matrix towards identity unless there is substantial {\em transfer error}. 
We show that, in the absence of any target domain labeled data or in case of zero transfer error, posterior means of class probability estimates from our approach coincide with those from the baseline learner, establishing that the naive estimation that ignores transfer error is a special case of our algorithm. We devise a novel, fast Gibbs sampler with augmented data for our Bayesian hierarchical model. 

We then extend our approach to one that uses an ensemble of input predictions from multiple classifiers. The ensemble model accomplishes method-averaging over different classifiers to reduce the risk of using one method that is inferior to others in a particular study. We establish a theoretical result that the class probability estimates from the ensemble model coincides with that from a classifier with zero transfer error. A Gibbs sampler for the ensemble model is also developed, as well as a computationally lighter version of the model that is much faster and involves fewer parameters. Simulation and data analyses demonstrate how the ensemble sampler consistently produces estimates similar to those produced by using our transfer learning on the single best classifier. 

Our approach is also post-hoc, i.e., only uses pre-trained baseline classifier(s), instead of attempting to retrain the classifier(s) multiple times with different versions of training data. This enables us to use publicly available implementations of these classifier(s) and circumvents iterative training runs of the baseline classifier(s) which can be time-consuming and inconvenient in epidemiological settings where data collection continues for many years, and the class probabilities needs to be updated continually with the addition of every new survey record. 

The rest of the manuscript is organized as follows. We present the motivating application in Section \ref{sec:motiv}. 
In Sections \ref{sec:cal} and \ref{sec:ens}, we present the methodology and its extension to the ensemble case. Section \ref{sec:map} presents an EM algorithm approach to obtain maximum a posteriori (MAP) estimates for the model, as a fast alternative to the fully Bayesian approach adopted earlier. Section \ref{sec:cov} considers the extension where class probabilities can be modeled as a function of covariates like age and sex, and spatial regions.  Section \ref{sec:sim} presents simulation results. Section \ref{sec:phmrc} returns to the motivating dataset and uses our transfer learning model to estimate national CSMFs for children deaths in India and Tanzania. We end the manuscript in Section \ref{sec:conc} with a discussion of limitations and future research opportunities. 

\subsection{Motivating dataset: }\label{sec:motiv} In low and middle income countries, it is infeasible to conduct full autopsies for the majority of deaths due to economic and infrastructural constraints, and/or religious or cultural prohibitions against autopsies \citep{abouzahr2015civil, allotey2015let}. An alternative method to infer the cause (or ``etiology") of death (COD) is to conduct {\it verbal autopsy} (VA) --  a systematic interview of the relatives of the deceased individual  -- to obtain information about symptoms observed prior to death \citep{soleman2006verbal}. 
Statisticians have developed several specialized classifiers that predict 
COD using the high-dimensional VA records as input. Examples include Tariff \citep{tariff, tariff2}, interVA \citep{interva}, insilicoVA \citep{insilico}, the King and Lu method \citep{king2008verbal}, 
expert alogirthm \citep{eava}, etc. The openVA R-package \citep{openva} has made many of these algorithms publicly available. Generic classifiers like random forests \citep{breiman2001random}, naive Bayes classifiers \citep{minsky1961steps} and support vector machines \citep{cortes1995support} have also been used \citep{flaxman2011random,miasnikof2015naive,koopman2015automatic}. Estimated COD labels for each VA record in a nationally representative VA database is aggregated to obtain national cause specific mortality fractions (CSMF) -- the population-level class membership probabilities, that are often the main quantities of interest for epidemiologists, local governments, and global health organizations. 

Formally, a CCVA algorithm is simply a classifier using the $S \times 1$ feature vector (VA report) $\bs$ to predict $c$ -- one of $C$ possible COD categories. Owing to the high-dimensionality of the feature space (VA record consists of responses to $200 - 350$ questions), learning this mapping $P(c \given \bs)$ requires substantial amount of {\em gold standard} (labeled) training data. 
Usually in the country of interest, VA records are available for a representative subset of the entire population, but gold standard cause of death (GS-COD) data are ascertained for only a very small fraction of these deaths. In other words, there is abundant unlabeled data but extremely limited labeled data in the target domain. Ongoing projects like the \href{https://www.jhsph.edu/research/centers-and-institutes/institute-for-international-programs/current-projects/countrywide-mortality-surveillance-for-action-comsa-in-mozambique/index.html}{Countrywide Mortality Surveillance for Action (COMSA)} in Mozambique and in Sierra Leone typify this circumstance, where, in addition to conducting a nationally representative VA survey, researchers will ascertain gold standard COD (GS-COD) for a small number of deaths from one or two local hospitals using minimally invasive autopsies (MIA) \citep{byass2016minimally}. Budgetary constraints and socio-cultural factors unfortunately imply that only a handful of deaths can eventually be autopsied (up to a few hundred).

\begin{figure}[h]\label{fig:mis}
	\centering
	\includegraphics[scale=0.47,trim={0 200 0 190},clip]{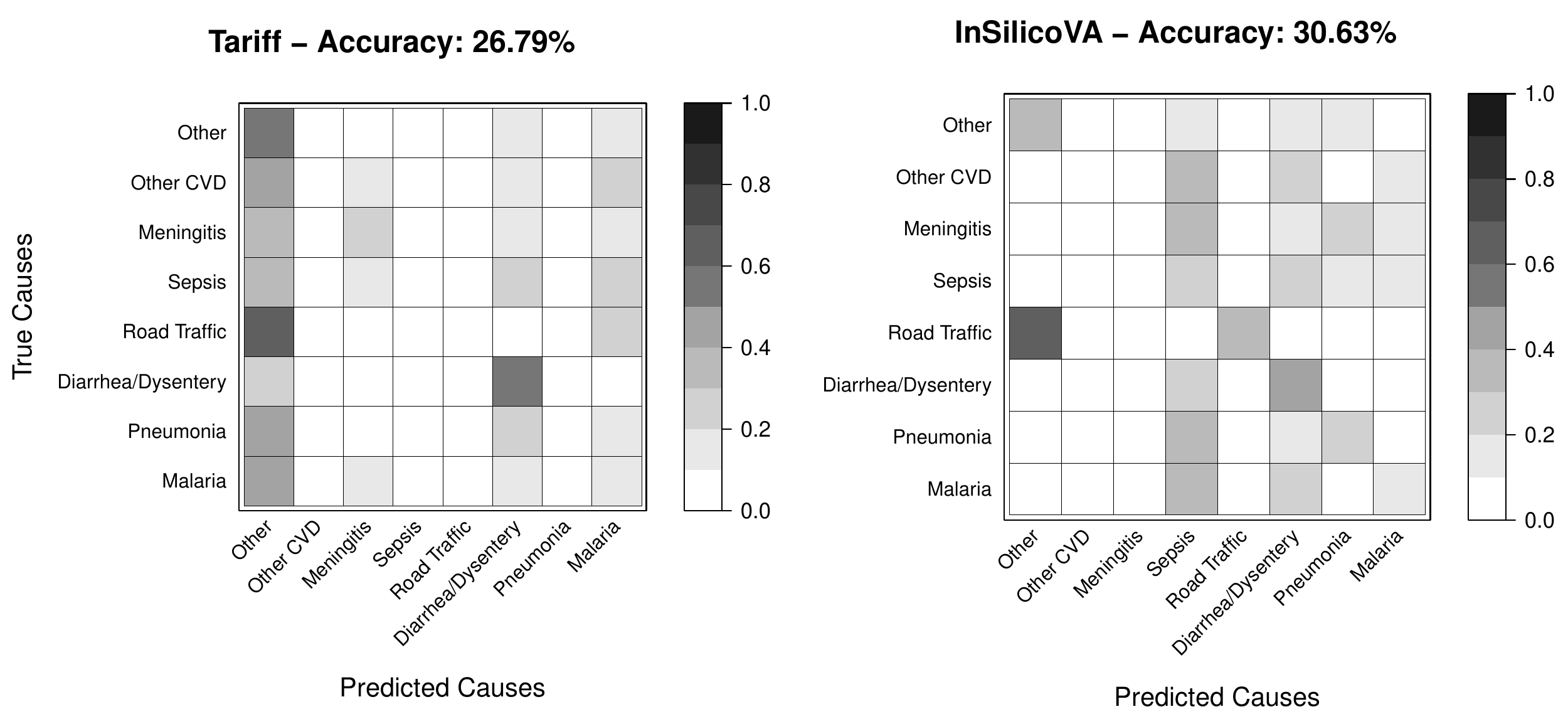}
	\caption{Misclassification matrices for PHMRC child cases in Tanzania using Tariff and InsilicoVA trained on all cases outside of Tanzania. CVD is abbreviation for cardio-vascular diseases.}
\end{figure}

Lack of sufficient numbers of labeled target-domain data implies that CCVA classifiers need to be trained on non-local source-domain data like the publicly available Population Health Metrics Research Consortium (PHMRC) Gold Standard VA database \citep{phmrc_murray}, that has more than $10,000$ paired physician and VA assessments of cause of death across 4 countries. 
However, there exists considerable skepticism about the utility of CCVA trained on non-local data as cause-symptom dynamics are often local in nature 
\citep{insilico,insilicoperformance}. 
To illustrate the issue, in Figure \ref{fig:mis}, we plot the misclassification matrices between the true COD of the PHMRC child cases in Tanzania against the predicted COD for these cases using two CCVA algorithms, Tariff and InsilicoVA, both trained on all PHMRC child data non-local to Tanzania. 
Both matrices reveal very large misclassification rates, some as high as $60\%$ or more. 
The large transfer errors indicate that the naive estimates of population-level class probabilities from CCVA classifiers trained on non-local source data are likely to be inaccurate thereby highlighting the need for transfer learning in this application. Additionally, like for any other application area, there exists considerable disagreement about which CCVA algorithm is the most accurate \citep{leitao2014comparison,insilico,insilicoperformance}. In our experience, no method is universally superior, and a robust ensemble transfer learning approach guarding against use of inaccurate classifiers is desirable. 
\begin{table}
\caption{\label{tab:gloss} Glossary of acronyms used in the manuscript}
	\centering
	\begin{footnotesize}
		\hskip -0.5in
		%\fbox{
		\begin{tabular}{c|c|c|c}
			Acronym & Full form & Acronym & Full form \\ \hline 
			VA & Verbal autopsy & PHMRC & Population Health Metrics Research Consortium \\
			CCVA & Computer coded VA & COMSA & Countrywide Mortality Surveillance for Action \\
			
			COD & Cause of Death & 	CSMF & Cause Specific Mortality Fraction \\
			CSMFA & CSMF accuracy & GS-COD & Gold-standard Cause of Death \\
			%CCC & Chance corrected concordance & MIA & Minimally Invasive Autopsy 
		\end{tabular}
	\end{footnotesize}
\end{table}

\section{Transfer learning for population-level class probabilities}\label{sec:cal}
\subsection{Naive approach}\label{sec:dim}
Let $\bp= (p_1,p_2,\ldots,p_C)'$ denote the true population-level class probabilities in a target domain $\calD_T$ where we have abundant unlabeled feature data, which we denote by $\calU$, and a very small labeled data $\calL$ of paired labels and features. For a feature vector $\bs$, we can write $p_i=P(G(\bs)=i)$ where $G$ denote the true (gold standard) class-membership.  
Also, let $A(\bs)=A(\bs \given \calG)$ denote the predicted class membership from the baseline classification algorithm $A$ trained on some large labeled dataset $\calG$ in a source domain $\calD_S$ different from $\calD_T$. If we do not use any transfer learning, the naive estimate of $\bp$ from $A$ is given by 
\begin{equation}\label{eq:base}
\widehat \bq = (\hat q_1,\ldots,\hat q_C)' \mbox{ where } \hat q_i = \sum_{\{s \in \calU\}} I(( A(\bs \given \calG)=i)) / N = v_i / N
\end{equation}
where $v_i$ is the number of instances in $\calU$ classified by $A$ to category $i$, and $N=\sum_i v_i$ is the sample size of $\calU$. If $\calU$ is large enough to be representative of the population in $\calD_T$, it is clear that 
\[\widehat \bq \approx \bq=(q_1,\ldots,q_C)' \mbox{ where } q_i = \int_\calU P(A(\bs \given \calG)  = i \given \bs) dP(\bs) = P(A(\bs) = i), \]
i.e., $\widehat \bq$ is the method-of-moments estimator of $\bq$. 

Unless the algorithm $A$ trained on $\calD_S$ perfectly agrees with the true membership assignment mechanism $G$ in $\calD_T$, there is no reason to consider $\bq$ or $\widehat \bq$ to be a good estimate of $\bp$. More realistically, since $\calD_S \neq \calD_T$, accuracy depends on how similar the algorithm $A$ is in the source and target domains. Hence, more generally we can think of $\bq$ as the expected population class probabilities in $\calD_T$ that would be predicted by $A(\cdot \given \calG)$. 

In their most general form, $G$ and $A$ can be thought of as measureable functions from the high-dimensional symptom space to the space of all $C$ dimensional simplexes. Hence, we can write 
\begin{equation}\label{eq:pop_csmf}
A(\bs) \sim \mult(\bq), \, G(\bs) \sim \mult(\bp)\, .
\end{equation}
This only depicts the marginal distributions of $A(\bs)$ and $G(\bs)$. To infer about $G$ from $A$, we need to model their joint distributions. We express
$q_j = \sum_{i=1}^C m_{ij} p_i$ where $m_{ij} = p (A(\bs) = j \given G(\bs) =i) $. In matrix notation, we have $\bq=\bM'\bp$ where $\bM=(m_{ij})$ is a transition matrix which we refer to as the {\em misclassification matrix}. First note that, if $\bM=\bI$, then $\bp=\bq$ and hence this subsumes the case where class probabilities from the baseline algorithm is trusted as reliable surrogates of the true class probabilities. 

For transfer learning to improve estimation of $\bp$, we can opt to use the more general relationship $\bq=\bM'\bp$ and estimate the misclassification rates
$m_{ij}$'s from $\calL$. Let $n = \sum_{i=1}^C n_i$ denote the sample size of $\calL$ with $n_i$ denoting the number of objects belonging to class $i$. Also let \[ \bT=(t_{ij})=(\sum_{\bs \in \calL} I(A(s)=j \given G(s)=i)) \] denote the {\em transfer error matrix} for algorithm $A$. 
Like many transfer learning algorithms, exploiting the transfer errors is key to our domain adaptation strategy. It is clear that $t_{ij}/n_i$ is a method-of-moments estimator of $m_{ij}$. 
In cases where there is substantial mismatch between the true and predicted categories for {\em instances} (samples) in $\calL$, we can use $\bT$ to estimate $\bM$ which along with the earlier estimate of $\bq$ can lead to a substantially improved estimate of $\bp$. Formally we can specify this via a hierarchical model as:
\begin{equation}\label{eq:calib}
\begin{array}{cc}
A(\bs_r) \iid& \mult (1, \bM'\bp), r=1,2,\ldots,N \\
\bT_{i*} \overset{ind}{\sim}& \mbox{Multinomial} (n_i,  \bM_{i*}), i=1,2,\ldots,C
\end{array}
\end{equation}
where for $r=1,2,\ldots,N$, $\bs_r$ denote the feature set for the $r^{th}$ instance in $\calU$, and for any matrix $\bM$, $\bM_{i*}$ and $\bM_{*j}$ denote its $i^{th}$ row and $j^{th}$ column respectively.  The top-row of (\ref{eq:calib}) represents the relationship $\bq = \bM'\bp$ and yields the method-of-moments estimators $\wh\bq = (v_1,v_2,\ldots,v_C)'/N$. 
The bottom-row of (\ref{eq:calib}) is consistent with the naive estimates $t_{ij}/n_i$ of $m_{ij}$. 

To estimate $\bp$, one can adopt a modular two-step approach where first $\wh\bq$ and $\wh\bM$ are calculated separately and then obtain
\[ \wh \bp = \underset{\bp: \bone'\bp =1,\; p_i \geq 0} {\arg \min} L(\wh \bq, \wh \bM'\bp)\]
where $L$ is some loss function like the squared-error or, more appropriately, the Kullback-Liebler divergence between the probability vectors. This approach fails to propagate the uncertainty in the estimation of $\bM$ in the final estimates of $\bp$. Benefits of a one-stage approach over a two-stage one has been demonstrated in recent work in transfer learning \citep{long2014adaptation}. We recommend  the one-stage information-theoretically optimal solution of estimating the joint MLE of $\bM$ and $\bp$ from (\ref{eq:calib}).

The advantage of this simple transfer learning method is that it circumvents the need to improve the individual predictions of $A$ in $\calD_T$, and directly calibrates the population-level class probabilities $\bp$, which are the quantities of interest here. We efficiently exploit the small local training data $\calL$ to reduce cross-domain bias. Instead of trying to use $\calL$ to estimate variants of a $S \times C$ matrix $\bP(\bs \given c)$ describing propensities of manifestation of each symptom given each cause, as is used by many CCVA algorithms like Tariff, InsilicoVA etc., we now only use $\calL$ to train a $p(A(\bs) \given c)$ misclassification matrix. One can view the predicted cause $A(\bs)$ as a one dimensional proxy feature. Consequently, the matrix $(P(A(\bs) \given c))$ involves only $C(C-1)$ parameters as opposed to the $\calO(SC)$ parameters of the $\bP(s \given c)$ matrix. For verbal autopsy data, $S$ is typically around $250$ while we can choose $C$ to be small focusing on the top $3-5$ causes. Hence, our approach achieves considerable dimension reduction by switching from the original feature space to the predicted class space.  

In equation \ref{eq:calib} above, $\bq$ can be estimated precisely because $N$ is large. However, $\bM$ has $C\times (C-1)$ parameters so that if there are many classes, the estimates of $m_{ij}$ will have large variances owing to the small size of $\calL$. Furthermore, in epidemiological studies, data collection often spans a few years; in the early stages, $\calL$ may only have a very small sample size resulting in an extremely imprecise estimate of $\bM$, even if we group the classes to a  handful of larger classes. Consequently, in the next section we propose a regularized approach that stabilizes the transfer learning.

\subsection{Bayesian regularized approach}\label{sec:bayes}
If $\calL$ was not available, i.e., there is no labeled data in the target domain, we only have $\calU$ and $\calG$. Then it would be natural to train $A$ using $\calG$ and predict on $\calU$ to obtain the estimates $\widehat \bq$ as the best guess for $\bp$. This is equivalent to setting $\bp = \bq$ and $\bM=\bI$, i.e., assuming that the algorithm $A$ perfectly classifies in $\calD_T$ even when trained only using $\calG$ from $\calD_S$. 
Extending this argument, when $\calL$ is very small, direct estimates of $\bM$ would be unstable and we should rely more on the predictions from $A$ trained on $\calD_S$. Hence, it is reasonable to shrink $\bp$ towards $\bq$ i.e., we shrink towards the default assumption that the baseline learner is accurate. This is equivalent to shrinking the estimate of $\bM$ towards $\bI$. The simplest way to achieve this is by using the regularized estimate $\widetilde \bM = (1- \lambda) \wh \bM + \lambda \bI$ where $\wh \bM = (\wh m_{ij}) = t_{ij} / n_i$ is the unshrunk method-of-moments estimate of $m_{ij}$ as derived in the previous section. The regularized estimate $\widetilde \bM$ (like $\wh\bM$ and $\bM$) remains a transition matrix. The parameter $\lambda$ quantifies the degree of shrinkage with $\lambda=0$ yielding the unbiased method-of-moments estimate and $\lambda=1$ leading to $\wh \bp = \wh \bq$. Hence, $\lambda$ represents the bias variance trade-off  for estimation of transition matrices and for small sample sizes some intermediate values of $\lambda$ may lead to better estimates of $\bM$ and $\bp$.

In epidemiological applications, as data will often come in batches over a period spanning few years, one needs to rerun the transfer learning procedure periodically to update the class probabilities. In the beginning, when $\calL$ is extremely small, it is expected that more regularization is required. Eventually, when $\calL$ becomes large, we could rely on the direct estimate $\wh \bM$. Hence, $\lambda$ should be a function of the size $n$ of $\calL$, with $\lambda = 1$ for $n=0$ and $\lambda \approx 0$ for large $n$. Furthermore, at intermediate stages, since the distribution of true class memberships in $\calL$ will be non-uniform across the classes, we will have a disparity in sample sizes $n_i$ for estimating the different rows of $\bM$. Consequently, it makes more sense to regularize each row of $\bM$ separately instead of using a single $\lambda$. A more flexible regularized estimate is given by
$\widetilde \bM_{i*} = (1 - \lambda_i) \wh \bM_{i*} + \lambda_i \bI_{i*}$. 
The row specific weights $\lambda_i$ should be chosen such that $\lambda_i = 1$ when $n_i = \sum_{j=1}^C t_{ij} = 0$, and $\lambda_i \approx 0$ when $n_i$ is large. One choice to accomplish this is given by $\lambda_i = \gamma_i / (n_i + \gamma_i)$ for some fixed positive $\gamma_i$'s. 

We now propose a hierarchical Bayesian formulation that accomplishes this regularized estimation of any transition matrix $\bM$. We consider a Dirichlet prior $\bM_{i*} \overset{ind}{\sim} \mbox{Dirichlet} (\gamma_i (\bI_{i*} + \eps \bone))$ for the rows of $\bM$. We first offer some heuristics expounding choice of this prior. We will have $\bM_{i*} \given  \bT_{i*}, \gamma_i \sim $ Dirichlet$(\bT_{i*} + \gamma_i (\bI_{i*} + \eps 1))$. Hence, \[E(\bM_{i*} \given  \bT_{i*}, \gamma_i) = \frac {\bT_{i*} + \gamma_i (\bI_{i*} + \eps \bone) }{n_i + \gamma_i (1 + C\epsilon)} \overset{\epsilon \rightarrow 0}{\rightarrow} (1 - \lambda_i) \frac {\bT_{i*}}{n_i}  + \lambda_i \bI_{i*} \mbox{ where } \lambda_i = \frac{\gamma_i}{n_i + \gamma_i} .\] Hence, using a small enough $\epsilon$, the Bayes estimator (posterior mean) for $\bM$ becomes equivalent with the desired shrinkage estimator $\widetilde \bM_{i*}$ proposed above. When $n=0$, the Bayes estimate $E(\bM \given \bT, \bgamma=(\gamma_1,\gamma_2, \ldots, \gamma_C)') \approx \bI$, and for large $n$, $E(\bM \given \bT, \bgamma)$ becomes the method-of-moments estimator $\wh\bM$. Hence, the Dirichlet prior ensures that in data-scarce setting, $\bM$ is shrunk towards $\bI$ and consequently $\bp$ towards $\bq$. We note that however this initial exposition for the posterior of $\bp$ are derived conditional on estimation of $\bM$ as an independent piece and ignores the data from $\calU$. In Theorem \ref{th:postp}, we will present a more formal result that looks at the properties of the marginal posterior of $\bp$. 

To complete the hierarchical formulation, we augment (\ref{eq:calib}) with the priors:
\begin{equation}\label{eq:hier}
\begin{array}{rl}
\bM_{i*} \overset{ind}{\sim}& \mbox{Dirichlet} (\gamma_i (\bI_{i*} + \eps \bone)), i=1,2,\ldots,C \\
\bp \sim& \mbox{Dirichlet} (\delta \bone) \\
\gamma_i \overset{ind}{\sim}& Gamma(\alpha,\beta), i=1,2,\ldots,C
\end{array}
\end{equation}

In practice, we need to use a small $\eps > 0$ to ensure a proper posterior for $\bM$ when any off-diagonal entries of $\bT$ are zero, which is very likely due to the limited size of $\calL$. Note that our model only uses the data from $\calL$ to estimate the conditional probabilities $P(A(\bs) \given G(\bs))$ for $\bs \in \calL$. We do not model the marginal distribution of $A(\bs)$ for $\bs \in \calL$ like we do for $\bs$ in $\calU$. This is because often data for the labeled set is collected under controlled settings, and marginal distribution of the features for the samples in $\calL$ is not representative of the true marginal distribution of the features in $\calD_T$. In our verbal autopsy application, the GS autopsies are conducted in one or two medical centers in a country and the marginal distribution of COD for deaths occurring in a hospital will possibly be very different from the countrywide CSMFs. Due to economic constraints, the number of deaths for which a GS autopsy procedure will be conducted is often predetermined based on the project budget. Hence, the deaths for GS autopsy often will be chosen in a balanced way to represent most of the major causes in the country. Further variation in the cause of death distribution for $\calL$ is introduced due to uncertainty of obtaining consent for conducting full autopsy. 
All of these virtually guarantee that even if $\calL$ consists of few hundred deaths with GS-COD, the marginal CSMF in $\calL$ is highly unlikely to represent the marginal CSMF in the population. Hence, we only use $\calL$ to estimate the conditional probabilities $\bM$. 

Our previous heuristic arguments, illustrating the shrinkage estimation of $\bM$ induced by the Dirichlet prior, are limited to the estimation of $\bM$ from $\calL$ as an independent piece and disregards the data and model for $\calU$, i.e. the first row of (\ref{eq:calib}). In a hierarchical setup, however, the models for $\calU$ and $\calL$ contribute jointly to the estimation of $\bM$ and $\bp$. We will now state a more general result that argues that for our full hierarchical model specified through (\ref{eq:calib}) and (\ref{eq:hier}), when there is no labeled data in $\calD_T$ or if the algorithm $A$ demonstrates perfect accuracy (zero transfer error) on $\calL$, then the marginal posterior estimates of $\bp$ from our model coincides with the  baseline estimates $\wh\bq$. Before stating the result, first note that the likelihood for $\ba=(A(\bs_1),A(\bs_2),\ldots,A(\bs_N))'$ can be represented using the sufficient statistics $\bv=(v_1,v_2,\ldots,v_C)'$. 
We can write 
$p(\ba) \propto \prod_{j=1}^C q_j^{v_j}$ and hence $\bp,\bM,\bgamma | \mbox{data} = \bp,\bM,\bgamma \given \bv, \bT$. 
\begin{theorem}\label{th:postp} If $\bT$ is a diagonal matrix, i.e., either there is no $\calL$, or $A$ classifies perfectly on $\calL$, then $\lim_{\epsilon \to 0} \bp \given  \bv, \bT \sim Dirichlet(\bv+\delta \bone)$. For $\delta=0$, $\lim_{\epsilon \to 0} E(\bp \given  \bv, \bT) = \widehat \bq$.
\end{theorem}
Note that Theorem \ref{th:postp} is a result about the posterior of our quantity of interest $\bp$, marginalizing out the other parameters $\bM$, and the $\gamma_i$'s from the hierarchical model specified through equations (\ref{eq:calib}) and (\ref{eq:hier}). We also highlight that this is not an asymptotic result and holds true for any sample size, as long as we choose $\eps$ and $\delta$ to be small. This is important as our manuscript pertains to epidemiological applications where $\calL$ will be extremely small and asymptotic results are not relevant. 

Theorem \ref{th:postp} also does not require any assumption about the underlying data generation scheme, and is simply a desirable property of our transfer learning model. If there is no labeled data in $\calD_T$, then it is natural to trust the $P(c \given \bs)$ map learnt by $A$ on a source domain and only learn the target domain marginal distributions of $\bs$ from $\calU$ to arrive at the estimates $\widehat \bq$ of $\bp$. 
Similarly, in the best case scenario, when $A$ is absolutely accurate for the target domain, Theorem \ref{th:postp} guarantees that our model automatically recognizes this accuracy and does not modify the baseline estimates $\widehat \bq$ from $A$. Additionally, the marginal posterior distribution for $\bp$ given by $Dirichlet(\bv)$ also nicely demonstrates the impact of the sample size $n$ of $\calL$. Since $Var(\bp \given data) \approx \calO(1/n)$  when $A$ is perfect on $\calL$, more samples in $\calL$ improves the posterior concentration of  $\bp$ around $\widehat \bq$. 
 The result of Theorem \ref{th:postp} is confirmed in simulations in Section \ref{sec:sim}. 

\subsection{Gibbs sampler using augmented data}\label{sec:gibbs}
We device an efficient implementation of the hierarchical transfer learning model using a data augmented Gibbs sampler. The joint posterior density can be expressed as $$
p(\bp, \bM, \bgamma \given  \bv, \bT) \propto \; p(\bv \given  \bM, \bp)p(\bT \given \bM)p(\bM \given \bgamma)p(\bp)p(\bgamma)$$ 
Let $\bp \given \cdot$ denote the full conditional distribution of $\bp$. We use similar notation for other full conditionals. First note that since $p(\bv \given  \bM, \bp) \propto \prod_j (\sum_i m_{ij}p_i)^{v_j}$, the full conditional densities $\bp \given \cdot$ and $\bM \given \cdot$ do not belong to any standard family of distributions, thereby prohibiting a direct Gibbs sampler. We here use a data augmentation scheme enabling a Gibbs sampler using conjugate distributions. 

The term $(\sum_{i} m_{ij} p_{i})^{v_j}$ can be expanded using the multinomial theorem, with each term corresponding to one of the partitions of $v_j$ into $C$ non-negative integers.

Equivalently we can write
$$(\sum_{i} m_{ij} p_{i})^{v_j} \propto E(\prod_{i} (m_{ij}p_i)^{b_{ij}}) \mbox{ where }  \bb_j =(b_{1j},\ldots,b_{Cj})'\sim \mbox{ Multinomial}(v_j,\bone/C).$$ 
Choosing $\bb_1, \bb_2, \ldots, \bb_C$ to be independent, we can express
$\prod_{j} (\sum_{i} m_{ij} p_{i})^{v_j} \propto  E(\prod_j \prod_i (m_{ij}p_i)^{b_{ij}} )$ where the proportionality constant only depends on the observed $v_j$'s. Using the augmented data matrix $\bB= (\bb_1, \bb_2, \ldots, \bb_C) = (b_{ij})$, we can write the complete posterior as
\begin{equation}\label{eq:postaug}
\begin{array}{cc}
p(\bp, \bM, \bgamma, \bB \given  \bv, \bT) \propto
& \prod_j \prod_i (m_{ij}p_i)^{b_{ij}} \times \prod_{i} p_{i}^{\delta - 1} \times \prod_i \gamma_i ^{\alpha-1} \exp(-\beta \gamma_i) \times \\ 
& \prod_{i} \left( \frac{\Gamma(C\gamma_i\epsilon + \gamma_i)}{\Gamma(\gamma_i\epsilon)^{C - 1}\Gamma(\gamma_i\epsilon + \gamma_i)} \prod_{j}(m_{ij})^{t_{ij} + \gamma_i\epsilon +  \gamma_i1(i  = j) - 1} \right)
\end{array}
\end{equation}
The full conditional distributions can be updated as follows (derivations omitted): 
\begin{align*}
\bb_j \given \cdot &\sim Multinomial(v_j, \frac 1{\sum_i M_{ij}p_i}( M_{1j}p_{1}, M_{2j}p_{2}, \ldots, M_{Cj}p_{C})')\\
\bM_{i*} \given  \cdot  &\sim Dirichlet(b_{i1} + \gamma_i\epsilon + t_{i1}, \ldots, b_{ii} + \gamma_i\epsilon + t_{ii} + \gamma_i, \ldots, b_{iC} + \gamma_i\epsilon + t_{iC}) \\
\bp \given  \cdot &\sim Dirichlet(\sum_{j} b_{1j} + \delta, \ldots, \sum_{j} b_{Cj} + \delta) \\
p(\gamma_i \given \cdot) &\propto \frac{\Gamma(C\gamma_i\epsilon + \gamma_i)}{\Gamma(\gamma_i\epsilon)^{C - 1}\Gamma(\gamma_i\epsilon + \gamma_i)} \gamma_i ^{\alpha-1} \exp(-\beta \gamma_i) \prod_{j}m_{ij}^{\gamma_i\epsilon +  \gamma_i1(i  = j) }
\end{align*}

The data augmentation ensures that, except the $C$ $\gamma_i$'s, which are updated using a metropolis random walk with log-normal proposals, all the other $\calO(C^2)$ parameters are update by sampling from standard distributions leading to an extremely fast and efficient Gibbs sampler. 

\section{Ensemble transfer learning}\label{sec:ens} 
Let there be $K$ classifiers $A^{(1)}, A^{(2)}, \ldots, A^{(K)}$ and let $\ba^{(k)}=(a_1^{(k)}, a_2^{(k)}, \ldots, a_N^{(k)})'$ be the predicted class memberships from the $k^{th}$ algorithm for all the $N$ instances in $\calU$. Let $\bv\kpow$ denote the vector of counts of predicted class memberships on $\calU$ using $A\kpow$. We expect variation among the predictions from the different classifiers and consequently among the baseline estimates of population-level class probabilities $\widehat \bq\kpow = \bv\kpow/N$ and their population equivalents $\bq\kpow=P(A\kpow(\bs))$. Since the true population class probability vector $\bp$ is unique, following Section \ref{sec:dim} we can write $\bq\kpow = (q_1\kpow,q_2\kpow,\ldots,q_C\kpow)' = \bM^{(k)'} \bp$ where $\bM\kpow = (m_{ij}\kpow)$ is now the classifier-specific misclassification matrix. The predicted class membership for the $r^{th}$ instance in $\calU$ by algorithm $A^{(k)}$, denoted by $a_r\kpow$, marginally follows a $\mult(1,\bq\kpow)$ distribution. We have $K$ such predictions for the same instance, one for each classifier, and these are expected to be correlated. So, we need to look at the joint distribution. A realistic assumption, to derive the joint distribution, is to assume that $a_r^{(1)}, a_r^{(2)}, \ldots, a_r\Kpow$ are independent conditional on $G(\bs_r)$, i.e. 
\begin{equation}\label{eq:condind}
p(a_r^{(1)}=j_1,a_r^{(2)}=j_2,\ldots,a_r^{(K)}=j_K \given G(\bs_r)=i) = \prod_{k=1}^K m_{ij_k}\kpow
\end{equation}
 Once again drawing the analogy that the prediction vector $(a_r^{(1)}, a_r^{(2)}, \ldots, a_r\Kpow)'$ is a lower-dimensional list of proxy features for the $r^{th}$ individual, this assumption is akin to the naive Bayes classifier assumption used to jointly model the probability of features given the true class membership. This implies that the marginal independence of the $a_r\kpow$'s will not generally hold. Instead we will have
\begin{equation}\label{eq:q}
p(\ba_r=\bj) = p(a_r^{(1)}=j_1,a_r^{(2)}=j_2,\ldots,a_r^{(K)}=j_K) = \sum_{i=1}^C \left(\prod_{k=1}^K m_{ij_k}\kpow\right) p_i = w_{\bj}
\end{equation} 
where $\bj=(\jall)$ denotes a $C\times 1$ vector index. 

From the limited labeled dataset $\calL$ in the target domain $\calD_T$, the classifier specific transfer error matrices $\bT\kpow = (t\kpow_{ij})=(\sum_{\bs \in \calL} I(A\kpow(\bs)=j \given G(\bs) = i))$ are also known and can be used  
to estimate the respective misclassification matrices $\bM\kpow$ in the same way $\bM$ was estimated from $\bT$ in Section \ref{sec:dim}. To introduce shrinkage in the estimation of $\bM\kpow$, like in Section \ref{sec:bayes}, we assign Dirichlet priors for each $\bM\kpow$ . 

Let $\bw$ denote a $C^K \times 1$ vector formed by stacking up all the $q_{\jall}$'s defined in (\ref{eq:q}). The full specifications for the ensemble model that incorporates the predictions from all the algorithms is given by:
\begin{equation}\label{eq:enshier}
\begin{array}{rl}
\ba_r \iid& \mult (1,\bw), r=1,2,\ldots,N \\
\bT_{i*}\kpow \overset{ind}{\sim}& \mbox{Multinomial} (n_i,  \bM\kpow_{i*}), i=1,2,\ldots,C;\; k=1,2,\ldots,K  \\
\bM\kpow_{i*} \overset{ind}{\sim}& \mbox{Dirichlet} (\gamma\kpow_i (\bI_{i*} + \eps \bone)), i=1,2,\ldots,C;\; k=1,2,\ldots,K  \\
\bp \sim& \mbox{Dirichlet} (\delta \bone) \\
\gamma\kpow_i \overset{ind}{\sim}& Gamma(\alpha,\beta), i=1,2,\ldots,C;\; k=1,2,\ldots,K 
\end{array}
\end{equation}

Although $\bw$ is a $C^K \times 1$ vector, courtesy of the conditional independence assumption (\ref{eq:condind}), it is only parameterized using the matrices $\bM\kpow$ and $\bp$ as specified in  (\ref{eq:q}), and hence involves $KC^2+C$ parameters. This ensures that there is adequate data to estimate the enhanced number of parameters for this ensemble method, as for each $\bM\kpow$ we observe the corrsponding trasfer error matrix $\bT\kpow$. The Gibbs sampler for (\ref{eq:enshier}) is provided in Section \ref{sec:ensgibbs} of the Supplement. To understand how the different classifiers are given importance based on their transfer errors on $\calL$, we present the following result:

\begin{theorem}\label{th:joint} If $\bT^{(1)}$ is diagonal with positive diagonal entries, and all entries of $\bT^{(k)}$ are $\geq 1$ for all $k>1$,  then $\bp \given data \sim \mbox{Dirichlet } (\bv^{(1)}+\delta)$. For $\delta=0$, $E(\bp \given data) = \bq^{(1)}$. 
\end{theorem}

Theorem \ref{th:joint} reveals that if one of the $K$ algorithms (which we assume to be the first algorithm without loss of generality) produce perfect prediction on $\calL$, then posterior mean estimate of $\bp$ from the ensemble model coincides with that of the baseline estimate from that classifier. 
The perfect agreement assumed in Theorem \ref{th:joint} will not occur in practice. However, simulation and data analyses will confirm that the estimate of $\bp$ from the ensemble model tend to agree with that from the single-classifier model in Section \ref{sec:bayes} with the more accurate algorithm. This offers a more efficient way to weight the multiple algorithms, yielding a unified estimate of class probabilities that is more robust to inclusion of an inaccurate algorithm in the decision making. In comparison, a simple average of estimated $\bp$'s from single-classifier transfer learning models for each of the $K$ algorithms would be more adversely affected by inaccurate algorithms.

\subsection{Independent ensemble model}\label{sec:ensind}
The likelihood for the top-row of (\ref{eq:enshier}) is proportional to $\prod_\bj w_\bj^{y_\bj}$ where $y_\bj=\sum_{\bs \in \calU} I(a^{(1)}=j_1,\ldots,a\Kpow(\bs)=j_K)$ denote the total number of instances in $\calU$ where the predicted class-memberships from the $K$ algorithms corresponds to the combination $\bj=(j_1,\ldots,j_K)'$. Even though $\calU$ will be moderately large (few thousand instances in most epidemiological applications), unless both $C$ and $K$ are very small ($C \leq 5$ and $K \leq 3$),  $y_\bj$'s will be zero for most of the $C^K$ possible combinations $\bj$. This will in-turn affect the estimates of $\bw$. For applications to verbal autopsy based estimation of population CSMFs, there are many CCVA algorithms (as introduced in Section \ref{sec:intro}), and researchers often want to use all of them in an analysis. We also may be interested in more than $3-5$ top causes. In such cases, the extremely sparse $C^K$ vector formed by stacking up the $\by_\bj$'s will destabilize the estimation of $\bw$. 
Also, the Gibbs sampler (see Section \ref{sec:ensgibbs}) of the joint-ensemble model introduces an additional $C^K$ independent multinomial variables of dimension $C$ thereby accruing substantial  computational overhead and entailing long runs of the high-dimensional Markov chain to achieve convergence. 

In this section, we offer a pragmatic alternative model for ensemble transfer learning that is computationally less demanding. From equation (\ref{eq:q}), we note that 
\begin{equation}
p(a_r\kpow = j_k) = \sum_{j_s : s \neq k} \sum_{i=1}^C \left(\prod_{k=1}^K m_{ij_k}\kpow\right) p_i = \sum_{i=1}^C m_{ij_k}\kpow p_i
\end{equation}
by exchanging the summations. Hence, the marginal distribution of $a_r\kpow$ is $\mult(1, \bq\kpow)$ where $\bq\kpow = (\bM\kpow) ' \bp$. This shows how the formulation of the ensemble model leads to the single-classifier model in Section \ref{sec:bayes} if we used only the predictions from the $k^{th}$ algorithm. It also allows us to model the $a_r\kpow$'s independently for each $k$, ignoring the correlation among the predictions in $\calU$ from the $K$ classifiers. We model as follows:
\begin{equation}\label{eq:ensind}
\ba_r=(a^{(1)}_r,a^{(2)}_r,\ldots,a\Kpow_r)' \iid \prod_{k=1}^K \mult (1, \bq\kpow), r=1,\ldots,N \\
\end{equation}
We replace the top-row of (\ref{eq:enshier}) with (\ref{eq:ensind}), keeping the other specification same as in (\ref{eq:enshier}). We call this the independent ensemble model. Note that, while we only uses the marginal distributions of the $a_r\kpow$'s ignoring their joint dependence, the joint distribution is preserved in the model for the transfer errors on $\calL$ specified in the second-row of (\ref{eq:enshier}), as all the $\bM\kpow$'s are tied to the common truth $\bp$ through the equations $\bq\kpow=\bM^{(k)'}\bp$. While the total number of parameters for the joint and independent ensemble models remain the same, eliminating the joint model for each of the $C^K$ combination of predicted causes from the $K$ algorithms allows decomposing the likelihood for (\ref{eq:ensind}) as product of individual likelihoods on $\calU$ for each of the $K$ classifiers. 
Additionally, the Gibbs sampler for the independent ensemble model is much simpler and closely resembles the sampler for the single-classifier model in Section \ref{sec:gibbs}. We only need to introduce $K$ $C\times C$ matrices $\bB\kpow=(\bb_1\kpow, \bb_2\kpow, \ldots, \bb_C\kpow)$,  one corresponding to each CCVA algorithm, akin to the matrix $\bB$ introduced in Section \ref{sec:gibbs}. 
The Gibbs sampler steps for the independent ensemble model are:
\begin{align*}
\bb_j\kpow \given \cdot &\sim Multinomial(v\kpow_j, \frac 1{\sum_i M\kpow_{ij}p_i}( M\kpow_{1j}p_{1}, M\kpow_{2j}p_{2}, \ldots, M\kpow_{Cj}p_{C})')\\
\bM\kpow_{i*} \given  \cdot  &\sim Dirichlet(\bB\kpow_{i*}+\bT\kpow_{i*}+\gamma\kpow_i \bI_{i*} +  \eps\bgamma\kpow_i \bone) \\
\bp \given  \cdot &\sim Dirichlet(\sum_k \sum_{j} b\kpow_{1j} + \delta, \ldots, \sum_k \sum_{j} b\kpow_{Cj} + \delta)
\end{align*}

Observe that the sampler for the independent model uses $CK$ additional parameters as opposed to $C^K$ parameters introduced in the joint sampler. This ensures that the MCMC dimensionality does not exponentially increase if more algorithms are included in the ensemble model. The theoretical result in Theorem \ref{th:joint} no longer remains true for the independent model. However, our simulation results in Section \ref{sec:simens} of the Supplement show that in practice it continues to put higher weights on the more accurate algorithm and consistently performs similar to or better than the joint model. 

\section{MAP estimation}\label{sec:map} So far, we have only discussed fully Bayesian implementations of the model in (\ref{eq:hier}). If full inferential output is superfluous and only posterior point-estimates of the parameters are desired, we outline a MAP (Maximum a posteriori) estimation for obtaining posterior modes of the parameters using an EM-algorithm. The data augmentation scheme introduced for the Gibbs sampler in \ref{sec:gibbs} is also seamlessly congruous with the EM algorithm.

In particular, we consider the vector $\bv$ and $\bT$ as the observed data and augment $\bB$ introduced in Section \ref{sec:gibbs} as the missing data to form the complete data likelihood $l(\bB,\bv,\bT \given \bM, \bp, \bgamma)$ which is proportional to (\ref{eq:postaug}). At the $s^{th}$ iteration, let $\bM^{[s]} = (m_{ij}^{[s]})$, $\bp^{[s]} = (p_i^{[s]})$ denote the current values of the parameters. Then 
\[ E^{[s]}(b_{ij} \given \bv, \bT) = \frac{v_j m^{[s]}_{ij}p_i^{[s]}}{\sum_i m^{[s]}_{ij}p_i^{[s]}} = \wh b_{ij}^{[s]},\] where $E^{[s]}$ denotes the expectation taken using the parameter values from the $s^{th}$ iteration.
The EM algorithm then proceeds as follows:

\begin{equation}\label{eq:estep}
\begin{array}{c}
\mbox{E-step: } E^{[s]}(\log l(\bB,\bv,\bT \given \bM, \bp, \bgamma) \given \bv,\bT) = 
\sum_i \Bigg(   \sum_j \Big( \wh b_{ij}^{[s]} \log (m_{ij}p_i)  + \\
(t_{ij}+\gamma_i\eps + \gamma_i I(i=j) -1)\log (m_{ij}) \Big) +   (\delta -1)\log p_i + h(\gamma_i)  \Bigg) 
\end{array}
\end{equation}
where $h(\gamma)= \log \left( \frac{\Gamma(C\gamma + \eps)}{\Gamma(\gamma \eps)^{C-1} \Gamma(\gamma \eps + \gamma)} \right) + (\alpha -1) \log \gamma - \beta \gamma$. Subsequently, the maximization step can be formulated as:
\begin{equation}\label{eq:mstep}
\mbox{M-step: } 
\begin{array}{cl}
m_{ij}^{[s+1]} = & \frac{\wh b^{[s]}_{ij} + t_{ij} + \gamma_i \eps + \gamma_i I(i=j) -1}{\sum_j (\wh b^{[s]}_{ij} + t_{ij}) + \gamma_i C \eps + \gamma_i  -C} \\
p_i^{[s+1]} = & \frac{\sum_j \wh b^{[s]}_{ij} + \delta -1}{\sum_i \sum_j \wh b^{[s]}_{ij} + C\delta  -C} \\
\gamma_i^{[s+1]} = & \arg \max_\gamma \sum_j (\gamma \eps + \gamma I(i=j) -1)\log (m_{ij}) + h(\gamma)
\end{array}
\end{equation}

The closed form expression of $\bM$ and $\bp$ in the M-step is a consequence of the data augmentation. This drastically accelerates the MAP estimation as we only need to conduct $C$ univariate optimizations, one corresponding to each $\gamma_i$. If instead the data augmentation was not exploited and only the observed likelihood was used, we would need to search an $O(C^2)$ dimensional space to obtain the MAP estimates. We can implement similar MAP estimation algorithms for the joint and independent ensemble models detailed in Section \ref{sec:ens}. We omit the steps here. 

\section{Demographic covariates and spatial information}\label{sec:cov}
The transfer-learning model introduced up to this point is focused on generating population-level  estimates of the CSMF $\bp$. An important extension for epidemiological applications would be to model $\bp$ as a function of covariates like geographic region, social economic status (SES), sex and age groups. This will enable the estimation of regional and age-sex stratified estimates, of interest for understanding the prevalence trends as well as guiding policy and allocation of resources aimed at improving public health. In this section, we generalize the model to accommodate covariates. We illustrate only for the single-classifier model in Section \ref{sec:bayes}; a similar approach extends the  ensemble model. 

Let $\bx_r$ denote a vector of covariates for the $r^{th}$ VA record in $\calU$. We propose the following modifications to the model for allowing covariate-specific class distributions $\bp_r = (p_{r1},p_{r2},\ldots,p_{rC})'$:
\begin{equation}\label{eq:hiercov}
\begin{array}{rl}
A(\bs_r) \ind& \mult (\bM'\bp_{r}), r=1,2,\ldots,N \\
p_{ri} =& \frac{exp(\bx_{r}'\bbeta_{i})}{\sum_{i=1}^{C}exp(\bx_{r}'\bbeta_{i})}, i=1, 2, \ldots, C, \bbeta_{C} = 0 \\
\bbeta_i \ind& N(\bm_{0i}, \bW_{0i}) 
\end{array}
\end{equation}
All other components of the original model in (\ref{eq:calib}) and (\ref{eq:hier}) remain unchanged. The middle row of (\ref{eq:hiercov}) specify a multi-logistic model for the class probabilities using the covariates. The top row uses the covariate specific $\bp_r$ to model the analogous class probabilities $\bq_r = \bM'\bp_r$ as would be predicted by $A$. Finally, the bottom row specifies Normal priors for the regression coefficients. The switch from a Dirichlet prior for $\bp$ to the multi-logistic model implies we can no longer directly leverage conjugacy in the Gibbs sampler. \cite{polson2013bayesian} proposed a Polya-Gamma data augmentation scheme to allow conjugate sampling for generalized linear models. We now show how our own data augmentation scheme introduced in Section \ref{sec:gibbs} harmonizes with the Polya-Gamma sampler to create a stramlined Gibbs sampler. 

\subsection{Gibbs sampler using Polya-Gamma scheme}\label{sec:gibbspol} 
We will assume there are $G$ unique combinations of covariate values -- for example, if there are four geographic regions and three age groups, then $G=12$. If we have a continuous covariate, then $G=N$, where $N$ is the number of subjects sampled in $\calU$. 
Then letting $g$, $g=1,\ldots,G$, represent a specific covariate combination $\bx_g$, we can again represent the likelihood for $\ba=(A(\bs_1),A(\bs_2),\ldots,A(\bs_N))'$ using the $G \times C$ sufficient statistics $\bV=(v_{gj})$
where $v_{gi}$ is the total number of subjects with covariate values $g$ that were predicted to have died of cause $i$. Let $\bbeta = (\bbeta_{1}, \bbeta_{2}, \ldots, \bbeta_{C-1})$. We now have
$$
p(\bV \given \bM, \bbeta) \propto \prod_{g=1}^{G}\prod_{j=1}^{C}(\sum_{i=1}^{C}m_{ij}p_{gi})^{v_{gj}}
$$
and the joint posterior density can now be expressed as $$
p(\boldsymbol{\beta}, \bM, \bgamma \given  \bV, \bT) \propto \; p(\bV \given  \bM, \boldsymbol{\beta})p(\bT \given \bM)p(\bM \given \bgamma)p(\boldsymbol{\beta})p(\bgamma)$$ 

The terms that are different from Section \ref{sec:gibbs} are $p(\bV \given  \bM, \boldsymbol{\beta})$ and $p(\boldsymbol{\beta})$. The sampling step for $\bgamma$ remains exactly the same as previously discussed. We will use a similar data augmentation strategy as in Section \ref{sec:gibbs} and combine with a Polya-Gamma data augmentation to sample from this posterior distribution. 
We expand the term $(\sum_{i}m_{ij}p_{gi})^{v_{gj}} \propto E(\prod_{i}(m_{ij}p_{gi})^{b_{gij}})$ where
$$
\bb_{gj} =(b_{g1j},\ldots,b_{gCj})'\ind \mbox{ Multinomial}(v_{gj},\bone/C).$$  
Let $\bB$ denote the $GC \times C$ matrix formed by stacking all the $\bb_{gj}$'s row-wise. We can write
$$
p(\boldsymbol{\beta}, \bB, \bM, \bgamma \given  \bV, \bT) \propto \; \prod_{g}\prod_{i}\prod_{j}(m_{ij}p_{gi})^{b_{gij}}\times p(\bT \given \bM)p(\bM \given \bgamma)p(\boldsymbol{\beta})p(\bgamma)
$$ 
The following updates ensue immediately:
\begin{align*}
\bb_{gj} \given \cdot &\sim Multinomial(v_{gj}, \frac 1{\sum_i M_{ij}p_{gi}}( M_{1j}p_{g1}, M_{2j}p_{g2}, \ldots, M_{Cj}p_{gC})')\\
\bM_{i*} \given  \cdot  & \sim Dirichlet \left(\bT_{i*} + \bgamma_i \bI_{i*} + \bgamma_i \bone + (\sum_{g}b_{gi1}, \ldots, \sum_{g}b_{giC})' \right) \\
\end{align*}
For $\bbeta_i$'s we introduce the Polya-Gamma variables $\omega_{gi}$'s and define $\bOmega_{i} = \text{diag}(\{\omega_{gi}\}_{g=1}^{G})$, $n_{g} = \sum_{j}v_{gj}$, and $\bkappa_i = (\kappa_{1i}, \ldots, \kappa_{Gi})'$ where $\kappa_{gi} = \sum_{j}b_{gij} - n_{g} / 2$. Defining $\bW_{i}^{-1} = \bX'\bOmega_{i}\bX + \bW_{0i}^{-1}$, we then have
\begin{align*}
\omega_{gi} \given \cdot &\sim PG(n_{g}, \bx_{g}^{T}\bbeta_{i} - c_{gi}) \mbox{ where } c_{gi}=\log(\sum_{k \neq i} exp(x_{g}^{T}\beta_{k})) \\
\bbeta_{i} \given \cdot &\sim \mathcal{N}(\bm_{i}, \bW_{i}) \mbox{ where } \bm_{i}= \bW_{i}\left(\bX'(\bkappa_{i} - \bOmega_{i}\bc_{i}) + \bW_{0i}^{-1}\bm_{0i}\right)
\end{align*}
Here $PG$ denotes the Polya-Gamma distribution and $\bc_i=(c_{1i},c_{2i},\ldots,c_{Gi})'$. This completes the steps of a Gibbs sampler where all the parameters except $\bgamma$ are updated via sampling from conjugate distributions. 
We can transform the posterior samples of $\boldsymbol{\beta}$ to obtain posterior samples of $p_{gi}$. Estimates of the marginal class distribution for the whole country can also be obtained by using the relationship $
p_{i} = \int p_{gi} d P(g)$ where an empirical estimate of the covariate distribution $P(g)$ can be obtained from $\calU$. 

\subsection{Covariate-specific transfer error} Note that in the model above, we have assumed that the transition matrix $\bM$ is independent of the covariates. We can also introduce covariates in modeling the conditional probabilities $m_{ij}$'s using a similar multi-logistic regression. This will be particularly useful if there is prior knowledge about covariate-dependent biases in the predictions from a classifier. We can then nicely incorporate this information into the model and the implementation will involve Polya-Gamma samplers for each row of $\bM$ in a manner exactly similar to the sampler outlined above (we omit the details). 

Since the local labeled data is limited, caution must be exercised while adopting this approach and one can only use a handful of covariates for modeling the misclassification rates. Alternatively, if such a shortlist of important covariates is not available, a broader set of covariates can be used in conjunction with Bayesian variable selection priors.

\section{Simulation studies}\label{sec:sim}
The Population Health Metrics Research Consortium (PHMRC) study, conducted in 4 countries across six sites, is a benchmark database of paired VA records and GS-COD of children, neonates and adults. PHMRC data is frequently used to assess performance of CCVA algorithms. 
We conduct a set of simulation studies using the PHMRC data to generate a wide range of plausible scenarios where the performance of of our transfer learning models needs to be assessed with respect to the popular CCVA algorithms. 
First, we randomly split the PHMRC child data ($2064$ samples) into three parts representing $\calG$, and initial $\calL$ and initial $\calU$ respectively using a 2:1:2 ratio, containing roughly $800$, $400$ and $800$ samples respectively. As accurate estimation of mortality fractions from most prevalent causes are usually the priority, we restrict our attention to four causes: the top three most prevalent causes in the target domain data ($\calL \cup \calU$) -- Pneumonia, Diarrhea/Dysentry,  Sepsis, and an {\em Other} cause grouping together all the remaining causes.  

We wanted to simulate scenarios where both a) the marginal distributions $P(c)=P(G(\bs)=c)$ of the classes, and b) the conditional distributions $P(c \given \bs)$ are different between the source and target domains. 
To ensure the latter, given a misclassification matrix $\bM = (m_{ij})$ we want $P(A(\bs)=j \given G(\bs)=i) = m_{ij}$ for any $\bs \in \calL \cup \calU$. 
We will achieve this by discarding the actual labels in $\calL \cup \calU$ and generating new labels  such that an algorithm $A$ trained on $\calG$ shows transfer error rates quantified by $\bM$ on $\calL \cup \calU$. Additionally, the new labels need to be assigned in a way to ensure that the target domain class probability vector is $\bp_\calU$, for any choice of $\bp_\calU$ different from the source domain class probabilities in $\bp_\calG$.

Note that if the true population class probabilities in $\calD_T$ needs to be $\bp_\calU$, then $\bq_\calU$, the population class probabilities as predicted by $A$ is given by $\bq_\calU = \bM' \bp_\calU$. Hence, we first use $A$ trained on $\calG$ to predict the labels for each $\bs$ in the initial $\calU$. We then resample $\bs$ from the initial $\calU$ to create a final $\calU$ such that the predicted labels of $A(\bs)$ has the marginal distribution $\bq_\calU$. 
Next, from Bayes theorem, 
\[ p(G(\bs) = i \given A(\bs) = j) = \frac{M_{ij}p_{\calU,i}}{\sum_i M_{ij}p_{\calU,i}} = \alpha_{ij}.\]
For $\bs$ in $\calU$ such that $A(\bs)=j$, we generate the new ``true" labels from $\mult(1,(\alpha_{1j}, \alpha_{2j},\ldots,\alpha_{Cj})')$. This data generation process ensures that for any $\bs$ in $\calU$ both $G(\bs) \sim \mult(1,\bp_\calU)$ and $A(\bs) \given G(\bs)=i  \sim \mult(1,\bM_{i*})$ are approximately true. We repeat the procedure for $\calL$, using the same $\bM$ but a different $\bp_\calL$. This reflects the reality for verbal autopsy data where the symptom-given-cause dynamics is same for all deaths $\calL \cup \calU$ in the new country, but the hospital distribution of causes $\bp_\calL$ is unlikely to match the population CSMF $\bp_\calU$. For resampling to create the final $\calL$, we also vary $n$ --- the size of $\calL$ as $50$, $100$, $200$ and $400$, to represent varying amount of local labeled that will be available at different stages of a project. 

We consider two choices of $A$: Tariff and InsilicoVA. For $\bM$, we use three choices -- a diagonal matrix $\bM_1 = \bI$ -- representing the case where the algorithm $A$ is perfect for predicting in the target domain, a matrix $\bM_2$ with two large off-diagonal entries and all other off-diagonal ones being zero -- representing the scenario where there are one or two systematic sources of bias in $A$ when trained on a source domain $\calD_S$ different from $\calD_T$, and $\bM_3$ which represents the scenario where there are many small misclassifications. Full specifications of the matrices are given in Section \ref{sec:simdetails} in the Supplement. 
To ensure that $\bp_\calU$ and $\bp_\calL$ are different, we generate pairs of probability vectors $(\bp_\calL, \bp_\calU)'$ from Dirichlet$(\bone)$ distribution and divide the cases into three scenarios:
{\em low:} CSMFA$(\bp_\calL,\bp_\calU) < 0.4$, {\em medium:} $0.4<$ CSMFA$(\bp_\calL,\bp_\calU) < 0.6$, and 
{\em high:} CSMFA$(\bp_\calL,\bp_\calU) > 0.6$.
Here CSMFA denoting the CSMF accuracy is a metric quantifying the distance of a probability vector ($\bp_\calL$) from a reference probability vector ($\bp_\calU$) and is given by \citep{murray2011robust}:
\[CSMFA(\bp_\calL, \bp_\calU) = 1 - \frac{||\bp_\calL -  \bp_\calU||_1 }{2\min \bp_\calU} .\] 
For each scenario, we generated $100$ pairs of $\bp_\calL$ and $\bp_\calU$. For each generated dataset, we use all the algorithms listed in Table \ref{tab:models} for predicting $\bp_\calU$.
\begin{table}
	\caption{\label{tab:models} List of models used to estimate population CSMF}
	\centering
	\hskip -0.4in \begin{tabular}{cc}
		Model name & Description \\ \hline
		Tariff$_\calG$ & Tariff trained on the source-domain gold standard data $\calG$ \\
		$\tc$ & Bayesian transfer learner using the output from Tariff$_\calG$ \\
		Insilico$_\calG$ & InsilicoVA trained on the source-domain gold standard data  $\calG$\\
		$\ic$ & Bayesian transfer learner using the output from Insilico$_\calG$ \\
		Ensemble$_{I}$ & Ensemble Bayesian transfer learner (independent) using Tariff$_\calG$ and Insilico$_\calG$ \\
	\end{tabular}	
\end{table}

\begin{figure}[]
	\centering
	\includegraphics[scale=0.75]{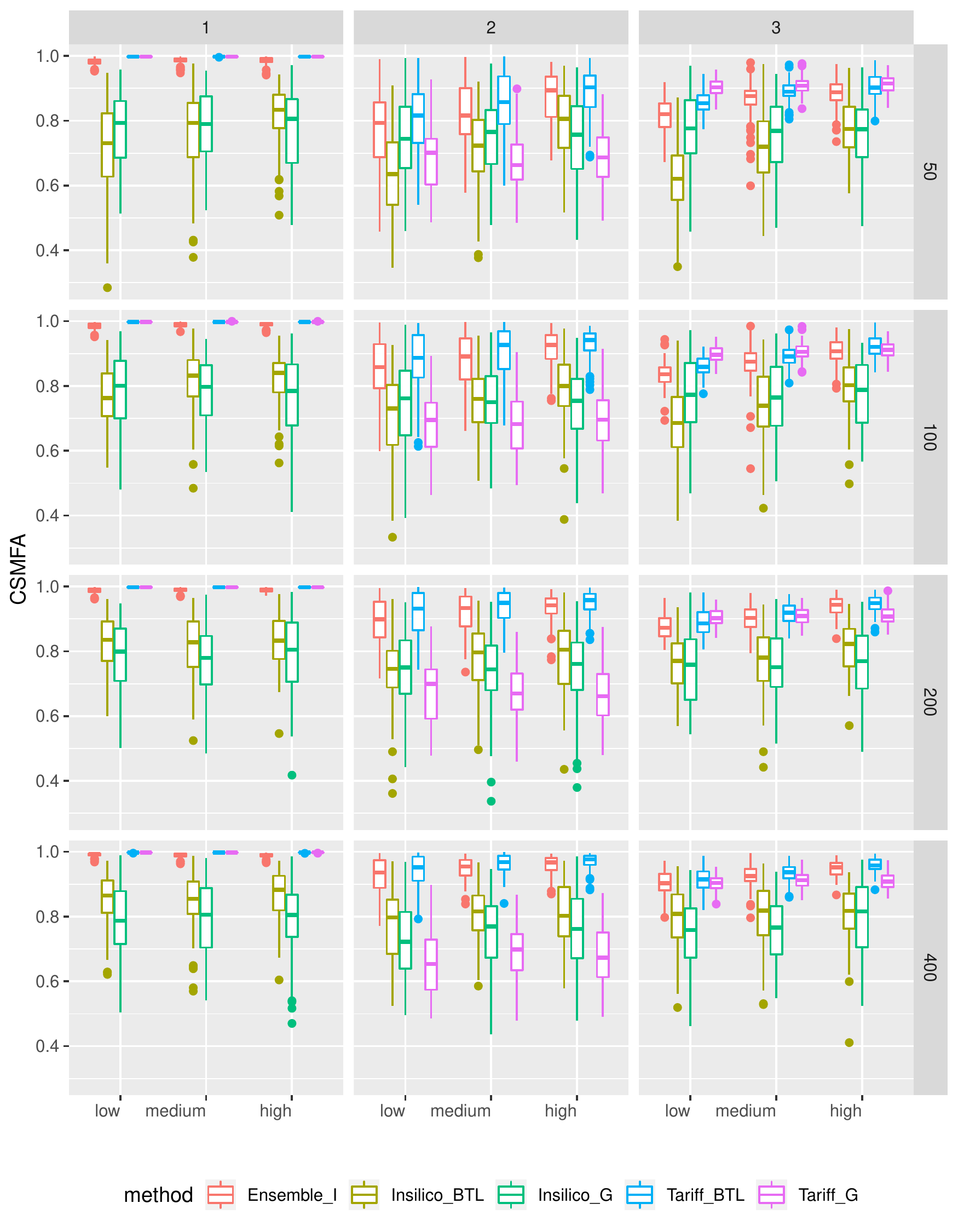}
	\vskip -7mm \caption{CSMF of ensemble and single-classifier transfer learners when data is generated using Tariff}\label{fig:csmfenstar}
\end{figure}

We present a brief summary of the results here. A much more detailed analysis is provided in Section \ref{sec:simdetails} of the Supplement. For an estimate $\wh \bp_\calU (x)$ generate by a model $x$, we assess the performance of $x$ using CSMFA$(x)$= CSMFA$(\wh \bp_\calU (x), \bp_\calU)$. Figure \ref{fig:csmfenstar} presents the CSMFA for all the five models for the case when the data was generated using Tariff via  the above design. 
The four rows correspond to four sample sizes of $\calL$ in increasing order, the three columns are for the three choices of $\bM$ described above, and in each figure the $x$-axis from left to right marks the {\em low, medium} and {\em high} settings. 

We first assess the performance of our Bayesian transfer learner as compared to the true baseline learner trained on $\calG$, 
i.e., we compare $\tg$ and $\tc$. We see that, first for $\bM_1$, there is no difference in CSMFA between the two methods. This corroborates the theoretical result in Theorem \ref{th:postp} as $\bM_1$ depicts the scenario where $A$ is perfect. For $\bM_3$, which depicts the situation that the CCVA algorithm is reasonably accurate, we see that when both the size of $\calL$ is small and when the CSMFA between $\calL$ and $\calU$ is low, i.e., there exists substantial disparity between the class distribution in $\calL$ and $\calU$, the  estimates from $\tc$ performs slightly worse than $\tg$. When either increases, $\tc$ performs better than $\tg$ with the improvements being substantial for the {\em high} setting and when $n$ is large. 
Finally, for $\bM_2$, which depicts large bias for $A$, we see that $\tc$ performs drastically better than $\tg$ analogue across all scenarios. Since it is impossible to gauge apriori the accuracy of the baseline learner, the results, across all three choices of $\bM$, provide strong support for the use of the domain-adapted VA for reducing large biases.

Next we turn our attention to $\ig$ and $\ic$. 
Since the data was generated using Tariff, both perform much worse than their Tariff analogues. We see the reverse trends in Figure \ref{fig:csmfensins} of the supplement where the data was generated using InsilicoVA. This shows how in applications some classifiers can perform poorly and even after transfer learning may still yield subpar performance compared to some other classifiers. This is where the value of the ensemble model is evident, as we see that the CSMFA for $\ei$, which uses the output from both Tariff and InsilicoVA, is consistently closer to that of $\tc$. In fact, for $\bM_1$, $\ei$ produces CSMFA identical to $\tc$ which is what Theorem \ref{th:joint} suggests. 
Similarly, for data generated using InsilicoVA Figure \ref{fig:csmfensins} demonstrates how the performance of $\ei$ aligns with that of $\ic$ which is the perfect classifier in that scenario. Since in practice we can never know which CCVA algorithm will be more accurate, the robustness offered by the ensemble sampler makes it a safer choice. 

In Section \ref{sec:simdetails} of the Supplement we present more thorough insights into the simulation study. In Section \ref{sec:simcsmf} we assess the impact of the disparity in the class distributions between the source and target domains. In Section \ref{sec:bias} we compare the biases in the estimates of individual class probabilities. Section \ref{sec:simsize} delves into the role of the sample size and quality of the limited labeled set $\calL$. Section \ref{sec:simnaive} demonstrates the value of the Bayesian shrinkage by comparing with the frequentist transfer learning outlined in Section \ref{sec:dim}. In Section \ref{sec:simens} we compare the joint and independent ensemble models and demonstrate how they favorably weight the more accurate algorithm. Finally, in Section \ref{sec:simcod}, we compare the performance of the models for predicting individual-level class probabilities for target domain data using the algortihm outlined in Section \ref{sec:cod}. 

\section{Predicting CSMF in India and Tanzania}\label{sec:phmrc}
We evaluate the performance of baseline CCVA algorithms and our transfer learning approach when predicting the CSMF for under 5 children in India and Tanzania using the PHMRC data with actual COD labels. We could take advantage of the fact that PHMRC was conducted in multiple countries and our strategy was to sample records from a given country into $\calL$ and $\calU$, while treating all records from outside of the country as the non local gold standard set ($\calG$). We used the PHMRC child data to specifically focus on cause distribution for children deaths. This is important as the aforementioned COMSA projects currently only aim to conduct minimally invasive autopsies for children deaths under the age of five. We used both India and Tanzania, as they were the only countries with substantial enough sample sizes ($N_{India} = 948$, $N_{Tanzania} = 728$).

For a given country (either India or Tanzania), we first split the PHMRC child data into subjects from within the country ($\calL \cup \calU)$ and from outside of the country ($\calG)$. We then randomly selected $n ( \in \{50, 100, 200, 400\}$)  subjects from within the country of interest to be in $\calL$, while the remaining subjects from within the country were put in $\calU$. We trained models $Insilico_{G}$ and $Tariff_{G}$ using the non local data $\calG$, which were then used to predict the top COD for all subjects in $\calL$ and $\calU$. All predicted causes of death that are outside of the top 3 COD within the country were changed to ``Other''. These predictions were then used to estimate the calibrated CSMF for $\tc$, $\ic$, and $Ensemble_{I}$. Since the true labels (GS-COD) are available in PHMRC, we calculated the true $\bp_\calU$ for a country as the empirical proportions of deaths from each cause, based on all the records within the country. 
This $\bp_\calU$ was used to calculate the CSMF accuracy of each of the models. This whole process was repeated $500$ times for each combination of country and value of $n$. This made sure that the results presented are average over $500$ different random splits of $\calL$ and $\calU$ for each country, and are not for an arbitrary split. 

\begin{figure}[t]
	\centering
	\includegraphics[scale=0.2]{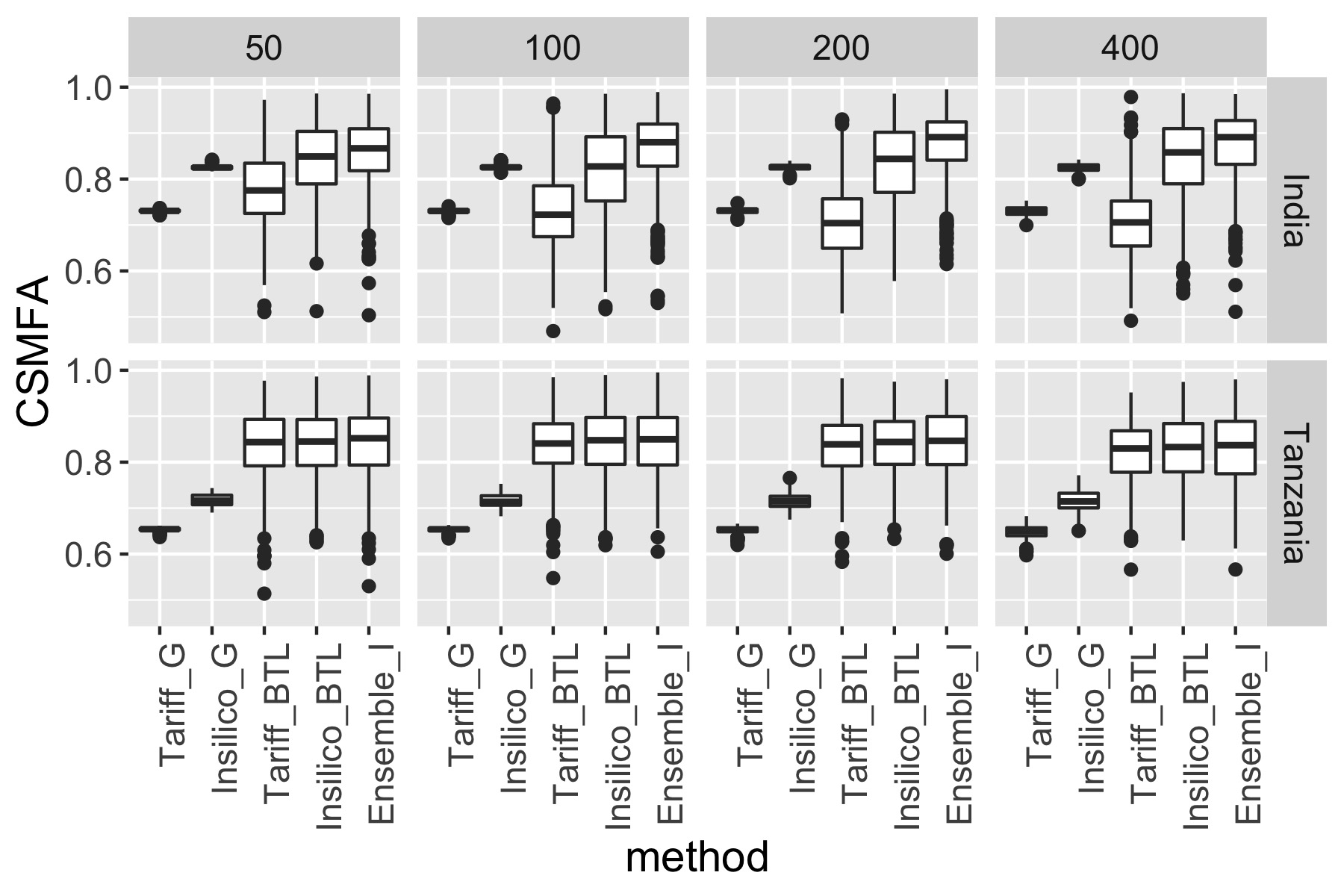}
	\vskip -7mm \caption{CSMFA using true GS-COD labels for the top 3 COD + others}\label{fig:realdata}
\end{figure}

Figure \ref{fig:realdata} presents the results of this analysis. 
The top and bottom rows represent the results for India and Tanzania respectively. The four columns correspond to four different choices of $n$.  
There are several notable observations. First, regardless of $n$, choice of algorithm $A$, and country, the calibrated estimates of prevalence from our transfer learning model performed better than or similar to the analogous baseline CSMFs, i.e., $\tc$ performed better than $\tg$, and $\ic$ performed better than $\ig$. Second, the magnitude of improvement for the our approach depends on the country and the size of $\calL$. 
Within India, the CSMFA of $\tc$ and $\ic$ is similar to respectively those from $\tg$ and $\ig$, with the InsilicoVA based learners performing significantly better. The ensemble model produced the highest accuracy, even though it used Tariff as one of the learners, which on its own was not performing well. 
In Tanzania, once again, the baseline InsilicoVA model $\ig$ does better than $\tg$ although the difference is less prominent than for India. For all choices of $n$, all the three transfer learning models, $\ic$, $\tc$ and the the ensemble model, produced substantially higher CSMFAs compared to those from the two baseline models. 

One caveat of this analysis is that we have fixed CSMF for the training and the test data, as determined empirically. Another downside is that, since we need to split the limited target-domain data into $\calL$ and $\calU$, as $n$ increases, the sample size $N$ of $\calU$ decreases. In fact, for Tanzania, when $n = 400$, there are more subjects in $\calL$ than in $\calU$. Thus the impact of increase in sample size $n$ for $\calL$ is confounded by the decrease in sample size $N$ for  $\calU$ which can make the estimation less precise. In Section \ref{sec:sens} of the Supplement we conduct a sensitivity analysis by resampling the records to create source and target-domain datasets which display varying degrees of concordance in the cause distribution, and where $\calU$ has a fixed sample size.

\section{Discussion}\label{sec:conc} 
Epidemiological studies pose unique challenges to transfer learning, stemming from its focus on estimating population-level quantities as opposed to individual predictions, small sample sizes coupled with high-dimensional feature spaces (survey records), and lack of large training databases typically available for many other machine learning tasks. Motivated by these settings, we have presented a parsimonious hierarchical model-based approach to transfer learning of population-level class probabilities, using abundant labeled source domain data, limited labeled target domain data, and abundant unlabeled target domain data. 

Shrinkage or regularization is at the core of our approach. In datasets with large numbers of variables (dimensions), regularized methods have become ubiquitous. A vast majority of the literature focuses on shrinking estimates (mostly regression coefficients and covariance or precision matrices) towards some known sub-model. We apply the same principle of regularization in a unique way for estimating the population class probabilities. Instead of shrinking towards any underlying assumptions about the true population distribution, we shrink towards the baseline estimate from a classifier trained on source data. In absence of sufficient target-domain data, this is the best available estimate and has to be used. We show how this shrinkage for the class probabilities is equivalent to shrinking the transfer error rate matrix towards the identity matrix. Subsequently, we show how properly chosen Dirichlet priors can achieve this shrinkage for this transition matrix. This regularized estimation of a transition matrix can also be applied in other contexts.

The fully Bayesian implementation is fast, owing to a novel data-augmented Gibbs sampler. The ensemble model ensures robust estimates via averaging over many classifiers and reduces the risk of selecting a poor one for a particular application.
Our simulations demonstrate the value of transfer learning, offering substantially improved accuracy. The PHMRC data analysis makes evident the value of collecting a limited number of labeled data GS-COD in the local population using full or minimally invasive autopsies, alongside the nationwide VA survey. Subsequently using transfer learning improves the CSMF estimates. The results also show how our approach benefits from larger sample sizes of the local labeled set $\calL$, and from closer alignment between the marginal class probabilities in $\calL$ and the true target domain class probabilities. 

For VA data, we note that while we have treated $\calG$ as a labeled dataset in the source domain $\calD_S$, in practice it can be any other form of gold standard information sufficient to train a VA classifier. CCVA methods like Tariff and the approach in \cite{king2008verbal} represent a traditional supervised learning approach and needs a substantial labeled training dataset $\calG$. InterVA is a semi-supervised learning approach where $\calG$ is a standard matrix of letter grades representing the propensity of each symptom given each cause. InsilicoVA generalizes InterVA and endows the problem with a proper probabilistic framework allowing coherent statistical inference. It adapts to the type of $\calG$ and can work with either the default symptom-cause matrix used in InterVA or estimate this matrix based on some labeled training data of paired VA and GS-COD records. Our transfer learning is completely agnostic to the choice of this baseline CCVA algorithm and the form of $\calG$ they require. We only need the predictions from that algorithm for all instances in $\calL \cup \calU$.

One important direction forward would be to generalize this approach for more complex COD outcomes. Currently COD outcome is viewed as a discrete variable taking values on a set of causes like Pneumonia or Sepsis. In practice, death is a complex chronological series of several events starting from some root causes and ending at the immediate or proximal cause. In addition to understanding prevalence of causes in the population, another goal for many of the aforementioned programs is to identify medical events that occurred before death for which an intervention could prevent or delay mortality. Extending the current setup for hierarchical or tree-structured COD outcome would be a useful tool to address this aim. Many CCVA algorithms, in addition to predicting the most likely COD, also predict the (posterior) distribution of likely causes. Our current implementation only uses the most likely COD as an input. An extension enabling the use of the full predictive distribution as an input can improve the method. In particular, this will benefit the individual COD predictions for which currently two individuals with the same predicted COD from CCVA have the same predicted COD distribution after transfer learning. Finally, the VA records, containing  about $250$ questions for thousands of individuals, naturally has several erroneous entries. Currently preprocessing VA records to eliminate absurd entries and records entails onerous manual labor. It is challenging to develop quality control models for VA data due to the high dimensionality of the symptoms. Akin to what we did here, one can consider dimension reduction via the predictions of CCVA algorithms to create an automated statistical quality control for VA records. 

%\appendix
%\section{Appendix: Glossary}\label{app:gloss}

%\pagebreak

%\bigskip
\begin{center}
	{\large\bf SUPPLEMENTARY MATERIAL}
\end{center}

\begin{description}
	
	\item[Supplement.pdf:] Contains proofs of Theorems \ref{th:postp} and \ref{th:joint}, transfer learning of individual class predictions, the Gibbs sampler for the joint ensemble model, detailed analysis of the simulated data, sensitivity analysis for the PHMRC data for India and Tanzania, and additional figures. 
	
	\item[R-package calibratedVA:] R-package `calibratedVA' containing code to obtain estimates of population CSMFs from our transfer learning approach using baseline predictions from any verbal autopsy algorithm is available at \url{https://github.com/jfiksel/CalibratedVA/}. The package also contains the code for the ensemble model for using outputs from several VA algorithms. A vignette describing how to navigate the package and demonstrating the use of the methodology is provided in \url{https://github.com/jfiksel/CalibratedVA/blob/master/vignettes/CalibratedVA.Rmd}
	
	%\item[HIV data set:] Data set used in the illustration of MYNEW method in Section~ 3.2. (.txt file)
	
\end{description}

\appendix

\section{Proofs}\label{app:proofs}

\begin{proof}[Theorem \ref{th:postp}] 
	The marginal posterior $\bp\given \bv, \bT$ is given by 
	$\int p(\bp, \bM, \bgamma \given  \bv, \bT) dP(\bM) dP(\bgamma)$. 
	Conditional on $\bgamma$, looking only at terms that involve $\bp$, $\bM$, and $\bgamma$, we have
	\begin{align*}
	p(\bp, \bM, \bv, \bT \given \bgamma)  \propto &\prod_{j} (\sum_{i} m_{ij} p_{i})^{v_j} \times \prod_{i} p_{i}^{\delta  - 1} \times \\
	&\prod_{i} \frac{\Gamma(\gamma_i(C\epsilon + 1))}{(\Gamma(\gamma_i\epsilon))^{C - 1}\Gamma(\gamma_i(\epsilon + 1))} \prod_{j}(m_{ij})^{t_{ij}+\gamma_i(\epsilon +  1(i  = j)) - 1}
	\end{align*}
	We will now use the multinomial theorem to expand the first product $\prod_{j} (\sum_{i} m_{ij} p_{i})^{v_j}$. Note that the $j^{th}$ term expands into $\kappa_j = {v_j + C - 1 \choose C - 1}$ terms, one corresponding to each partition of $v_j$. 
	Let $\bB^{(j)} = (b^{(j)}_{k i})$ denote the $\kappa_j \times C$ partition matrix formed by stacking up all $1\times C$ rows that represent a non-negative integer partition of $v_j$. The $k^{th}$ row of $\bB^{(j)}$ gives the $k^{th}$ partition and $i^{th}$ element of that row corresponds to power index for the $i^{th}$ term ($m_{ij}p_i$).
	We now have,
	\begin{align*}
	p(\bp, \bM \given \bv, \bT, \bgamma) \propto & \left(\prod_{j}\sum_{k_j = 1}^{\kappa_j}  \prod_{i} \frac{(m_{ij}p_{i})^{b^{(j)}_{k_ji}}}{b^{(j)}_{k_ji} !} \right) \times \prod_{i} p_{i}^{\delta  - 1} \times \\
	&\prod_{i} \frac{\Gamma(\gamma_i(C\epsilon + 1))}{(\Gamma(\gamma_i\epsilon))^{C - 1}\Gamma(\gamma_i(\epsilon + 1))} \prod_{j}(m_{ij})^{t_{ij}+\gamma_i(\epsilon +  1(i  = j)) - 1} \\
	\propto 
	&\sum_{k_1 =1}^{n_1} \cdots \sum_{k_C =1}^{n_C} \left( \prod_{i} \frac{\Gamma(\gamma_i(C\epsilon + 1)) p_{i}^{\sum_j b^{(j)}_{k_ji}-1}}{(\Gamma(\gamma_i\epsilon))^{C - 1}\Gamma(\gamma_i(\epsilon + 1))} \times \right. \\
	& \left. \prod_{j} \frac{ (m_{ij})^{b^{(j)}_{k_ji} + t_{ij} + \gamma_i(\epsilon + 1(i  = j)) - 1}}{b^{(j)}_{k_ji} !} \right) 
	\end{align*}
	
	Given $k_1, \ldots, k_C$ and $i$, the product $\prod_{j=1}^{C} (m_{ij})^{b^{(j)}_{k_ji} + t_{ij}+\gamma_i(\epsilon + 1(i  = j)) - 1}$ is the kernel of a $Dirichlet(b^{(1)}_{k_1i} + t_{i1}+ \gamma_i \epsilon, \ldots, b^{(i)}_{k_ii} +  t_{ii}+\gamma_i (\epsilon + 1), \ldots, b^{(C)}_{k_Ci} + t_{iC}+ \gamma_i \epsilon)$ distribution. Hence,  integrating $\bM$ out with respect to the order $\prod_{i=1}^{C}\prod_{j=1}^{C} dm_{ij}$, we are left with 
	
	$$
	p(\bp \given \bv, \bT, \bgamma) \propto \sum_{k_1 =1}^{n_1} \cdots \sum_{k_C =1}^{n_C} w_{k_1,k_2,\ldots,k_C}(\bgamma,\eps)  \prod_{i} p_{i}^{\sum_{j} b^{(j)}_{k_ji}+\delta -1}
	$$
	where $w_{k_1,k_2,\ldots,k_C}(\bgamma,\eps) = \prod_{i} \frac{\Gamma(\gamma_i(C\epsilon + 1)) \prod_{j=1}^{C} \Gamma(b^{(j)}_{k_ji} +  t_{ij}+ \gamma_i(\epsilon + 1(i=j)))}{(\Gamma(\gamma_i\epsilon))^{C - 1}\Gamma(\gamma_i(\epsilon + 1)) \Gamma( \sum_{j} (b^{(j)}_{k_ji} +  t_{ij})+\gamma_i (C \epsilon + 1))\prod_j b^{(j)}_{k_ji} !}$. 
	Hence, 
	\begin{align*}
	\bp \given \bv, \bT \sim \sum_{k_1 =1}^{n_1} \cdots \sum_{k_C =1}^{n_C} \left( \left( \int \frac 1{W(\bgamma,\eps)}w_{k_1,k_2,\ldots,k_C}(\bgamma,\eps) dF(\bgamma) \right) \times \right. \\
	\left. Dirichlet(\sum_{j} b^{(j)}_{k_j1}+\delta,\ldots,\sum_{j} b^{(j)}_{k_jC}+\delta) \right)
	\end{align*}
	where $W(\bgamma,\eps)=\sum_{k_1 =1}^{n_1} \cdots \sum_{k_C =1}^{n_C} w_{k_1,k_2,\ldots,k_C}(\bgamma,\eps)$. Without loss of generality, let the first row of each $\bB^{(j)}$ represent the partition of $v_j$ which allocates $v_j$ to the $j^{th}$ component and $0$ to all the other components. 
	For any $(k_1,k_2,\ldots,k_C)' \neq \bone_C$, we have 
	\begin{align*}
	\underset{\eps \rightarrow 0}\lim  \frac {w_{k_1,k_2,\ldots,k_C}(\bgamma,\eps) } {w_{1,1,\ldots,1}(\bgamma,\eps) } = \prod_i \left( \frac{\Gamma( \sum_{j} t_{ij}+v_i+\gamma_i) \Gamma(b^{(i)}_{k_ii} +  t_{ii}+ \gamma_i)}{\Gamma( \sum_{j} (b^{(j)}_{k_ji} +  t_{ij})+\gamma_i ) \Gamma(v_i +  t_{ii}+ \gamma_i)} \times \right. \\
	\left. \left(  \prod_{j \neq i} \underset{\eps \rightarrow 0}\lim \frac{\Gamma(b^{(j)}_{k_ji} +  t_{ij}+ \gamma_i \epsilon )}{\Gamma( t_{ij}+ \gamma_i \epsilon)} \right) \right)
	\end{align*}
	If $b^{(j)}_{k_ji} = 0$, the ratio $\frac{\Gamma(b^{(j)}_{k_ji} +  t_{ij}+ \gamma_i \epsilon )}{\Gamma( t_{ij}+ \gamma_i \epsilon)}$ is one. However, since $(k_1,k_2,\ldots,k_C)' \neq \bone_C$, we have atleast one pair $i \neq j$ such that $b^{(j)}_{k_ji} \geq 1$ and consequently 
	$$\frac{\Gamma(b^{(j)}_{k_ji} +  t_{ij}+ \gamma_i \epsilon )}{\Gamma( t_{ij}+ \gamma_i \epsilon)} = \prod_{s=0}^{b^{(j)}_{k_ji} - 1}(s+t_{ij}+\gamma_i \eps) \overset{\eps \rightarrow 0}\longrightarrow 0$$
	since $T$ is diagonal. Hence, $w_{1,1,\ldots,1}$ dominates all the other weights in the limiting case. Since each of the scaled weights are less than one, using dominated convergence theorem, 
	$$ \underset{\eps \rightarrow 0}\lim \;\int \frac 1{W(\bgamma,\eps)}w_{k_1,k_2,\ldots,k_C}(\bgamma,\eps) dF(\bgamma)= 1((k_1,k_2,\ldots,k_C)' = \bone)$$ and hence 
	$ \underset{\eps \rightarrow 0}\lim \; p(\bp \given \bv, \bT) \propto \prod_i p_{i}^{\sum_{j} b^{(j)}_{1i}+\delta -1} = \prod_i p_{i}^{v_i+\delta -1}  $.
\end{proof}

\begin{proof}[Theorem \ref{th:joint}]
	We proof only for the case $K=2$ as the same proof generalizes for arbitrary $K$. We simplify the notation for the proof. Let $v_{st}$ denote the number of instances in $\calU$ assigned to class $s$ by algorithm $1$, and class $t$ by algorithm $2$. We write $\bM\opow=\bM$, $\bM\tpow=\bN$, $\bT\opow=\bT$ and $\bT\tpow=\bU$ to get rid of the superscripts. Also, let $\bB_{(st)} = (b_{li}^{(st)})$ denote a $\kappa_{st} \times C$ matrix formed by stacking row-wise all possible partitions of $v_{st}$ into $C$ non-negative integers. Here $\kappa_{st} = {v_{st} +C-1 \choose C -1 }$ denotes the total number of such partitions. Let $\bh=(h_{11},h_{12},\ldots,h_{CC})'$ denote a generic index vector such that each $h_{st} \in \{1,2,\ldots,\kappa_{st}\}$ indexes a partition of $v_{st}$ and $\calH$ denote the collection of all such $\bh$'s. Then likelihood for $(\ba_1,\ba_2,\ldots,\ba_N)'$ is
	\begin{align*} \prod_{s=1}^C \prod_{t=1}^C \left(\sum_{l=1}^{\kappa_{st}} \prod_{i=1}^C \frac {(p_i m_{is} n_{it} )^{b^{(st)}_{li}}}{b^{(st)}_{li} !}  \right) &=\sum_{\bh \in \calH} \prod_{i=1}^C  \prod_{s=1}^C \prod_{t=1}^C \frac {(p_i m_{is} n_{it} )^{b^{(st)}_{h_{st}i}}}{b^{(st)}_{h_{st}i} !} \\
	&= \sum_{\bh \in \calH} \left(  \frac1{c_\bh}\prod_{i=1}^C p_i^{\sum_{s=1}^C \sum_{t=1}^C b^{(st)}_{h_{st}i}} \times \right.\\
	& \quad \left. \prod_{s=1}^C m_{is}^{\sum_{t=1}^C b^{(st)}_{h_{st}i}}\prod_{t=1}^C n_{it}^{\sum_{s=1}^C b^{(st)}_{h_{st}i}} \right)
	\end{align*}
	where $c_\bh$ is a constant term free of the parameters. 
	
	Incorporating the priors and marginalizing with respect to $\bM$ and $\bN$ we have
	
	\begin{align*}
	p(\bp \given \bv^*, \bT, \bU, \bgamma^{(1)}, \bgamma^{(2)}) \propto & \sum_{\bh \in \calH} \frac 1 {c_\bh} \prod_{i=1}^C \Bigg( p_i^{\sum_{s=1}^C \sum_{t=1}^C b^{(st)}_{h_{st}i} + \delta -1 } \times \\
	& \frac {\prod_{s=1}^C \Gamma(\sum_{t=1}^C b^{(st)}_{h_{st}i} + t_{is}+ \gamma_i^{(1)}(\eps+I(i=s))}{\Gamma\left(\sum_{s=1}^C \left(\sum_{t=1}^C b^{(st)}_{h_{st}i} + t_{is}\right) + \gamma_i^{(1)}(C\eps+1)\right)} \times \\
	& \frac {\prod_{t=1}^C \Gamma(\sum_{s=1}^C b^{(st)}_{h_{st}i} + u_{it}+ \gamma_i^{(2)}(\eps+I(i=t))}{\Gamma\left(\sum_{t=1}^C \left(\sum_{s=1}^C b^{(st)}_{h_{st}i} + u_{it}\right) + \gamma_i^{(2)}(C\eps+1)\right)} \Bigg) \\
	& \propto \sum_{\bh \in \calH}  w_\bh(\gamma^{(1)},\bgamma^{(2)},\eps) \; \prod_{i=1}^C p_i^{\sum_{s=1}^C \sum_{t=1}^C b^{(st)}_{h_{st}i} + \delta -1 } 
	\end{align*}
	where $w_\bh(\gamma^{(1)},\bgamma^{(2)},\eps)$ is the weight comprising of all the terms not involving $p_i$'s. Now, let $\calH^*$ denote the subset of $\calH$ such that for all $\bh^* = (h_{11}^*,h_{12}^*,\ldots,h_{CC}^*)' \in \calH^*$, each index $h_{st}^*$ corresponds to a partition of $v_{st}$ which allocates $v_{st}$ to the $s^{th}$ partition and zero to all the other partitions. Clearly, for any $\bh^*$, $\sum_{s=1}^C \sum_{t=1}^C b^{(st)}_{h^*_{st}i} = \sum_{s=1}^C \sum_{t=1}^C v_{st} I(i=s) = \sum_{s=1}^C I(i=s) v_s = v_i$.
	
	Let $\zeta$ denote a generic positive constant which does not depend on $\epsilon$. We absorb terms of the form $\lim_{\eps \rightarrow 0} \Gamma(x+\calO(\eps))$ where $x$ is always greater than $1$ into $\zeta$, as these limits will be non-zero. Noting that $t_{is}=0$ if $s \neq i$, for any $\bh^* \in \calH$ and $\bh \in \calH \setminus \calH^*$, we have
	\begin{align*}
	\underset{\eps \rightarrow 0}\lim  \; \frac {w_\bh}{w_\bh^*} = \zeta \prod_{i=1}^C \frac {\prod_{s \neq i} \Gamma(\sum_{t=1}^C b^{(st)}_{h_{st}i} + \gamma_i^{(1)} \eps)}{\prod_{s \neq i} \Gamma(\gamma_i^{(1)} \eps)} \frac {\prod_{t \neq i} \Gamma(\sum_{s=1}^C b^{(st)}_{h_{st}i} + u_{it}+ \gamma_i^{(2)} \eps)}{\prod_{t \neq i} \Gamma(\sum_{s=1}^C b^{(st)}_{h^*_{st}i} + u_{it}+ \gamma_i^{(2)} \eps)}
	\end{align*}
	Since $u_{it}$'s are greater than zero and there exists at least one pair $(i,s)$ such that $\sum_{t=1}^C b^{(st)}_{h_{st}i} > 0$, the result follows. 
\end{proof}

\section{Individual-level transfer learning}\label{sec:cod}
While our Bayesian transfer learning is primarily targeted to estimate population-level class probabilities, it can also be used to predict individual-level class probabilities in the target domain $\calD_T$. 
The posterior distribution of the true class membership $G(\bs_r)$ of the $r^{th}$ individual is given by
\begin{align*}
p(G(\bs_r) =i \given \ba, \bT) =& \int p(G(\bs_r) =i \given \bp, \bM, \bgamma, \ba, \bT) \times \\
& \quad p(\bp, \bM, \bgamma \given \bv, \bT)\; dP(\bM)\; dP(\bp)\; dP(\bgamma) \nonumber \\
=& \int p(G(\bs_r) =i \given \bp, \bM, A(\bs_r)) p(\bp, \bM \given \bv, \bT)\; dP(\bM)\; dP(\bp)\;\\ 
=& \int \frac{m_{ij}p_i}{\sum_{i=1}^C m_{ij}p_i} p(\bp, \bM  \given , \bT)\; dP(\bM)\; dP(\bp) \nonumber
\end{align*}
We can now easily conduct composition sampling using posterior samples of $\bM$ and $\bp$ to generate a posterior distribution for $G(\bs_r)$. This simple application of the Bayes theorem, can recover the individual class memberships. However, it is a crude approach because the posterior distribution of the $G(\bs_r)$ are identical for all instances $\bs_r$ with the same predicted class $A(\bs_r)$ from $A$. If $A$ is a probabilistic classifier like InsilicoVA \citep{insilico}, then in addition to providing a predicted class membership $A(\bs_r)$, $A$ also provides the predicted distribution for each individual's class. Utilizing the entire predicted distribution from $A$ should lead to improved individual level transfer learning. Since the focus of this manuscript is population level transfer learning, we do not further explore this avenue here. 

\section{Gibbs sampler for the joint ensemble model}\label{sec:ensgibbs} Let $y_{\bj}$ be the number of instances in $\calU$ for which algorithm $A^{(1)}$ predicts cause $j_1$, $A^{(2)}$ predicts cause $j_2$, and so on. Let $\by^*$ be the $C^K \times 1$ vector formed by stacking the $y_\bj$'s. Also, let $u_{i\bj}=\prod_{k=1}^K m_{ij_k}\kpow$ and $\bu_{\bj} = (u_{1\bj}, u_{2\bj}, \ldots, u_{C\bj})'$. 

The posterior $\bp, \{\bM^{(k)},\bgamma_k\}_{k=1,\ldots,K} \given \bT^{(1)}, \bT^{(2)}, \ldots, \bT\kpow, \bw^*$ is proportional to
\begin{align*}
& \prod_{\bj} (\sum_{i} u_{i\bj} p_{i})^{y_{\bj}} \times \prod_{i} p_{i}^{\delta  - 1} \times \\
\prod_{k=1}^K & \left( \prod_{i=1}^C \frac{\Gamma(\gamma\kpow_i(C\epsilon + 1))}{(\Gamma(\gamma\kpow_i\epsilon))^{C - 1}\Gamma(\gamma\kpow_i(\epsilon + 1))} \times \right. 
\left. \prod_{j}(m\kpow_{ij})^{t_{ij}\kpow+\gamma\kpow_i(\epsilon +  1(i  = j)) - 1} \right).
\end{align*}
We will once again use data augmentation to implement the Gibbs sampler. Let $\bb_{\bj} = (b_{1\bj}, b_{2\bj}, \ldots, b_{C\bj})')$ denote the $C \times 1$ dimensional realization of a Multinomial $(y_{\bj}, \bone/C)$ distribution, and let $\bB$ denote the $C^K \times C$ matrix formed by stacking the independent $\bb_{\bj}$'s row-wise for all combinations of $\bj$. Then we have the following full conditionals for the Gibbs sampler:
\begin{align*}
\bb_\bj \given \cdot &\sim Multinomial(y_\bj, \frac 1{\bone' (\bu_{\bj} \odot \bp)} \bu_{\bj} \odot \bp)\\
\bM\kpow_{i*} \given  \cdot  &\sim Dirichlet \left(\bT_{i*} + \bgamma_i\kpow \bI_{i*} + \bgamma_i\kpow \bone + (\sum_{\bj: j_k=1} b_{i\bj}, \ldots, \sum_{\bj: j_k=C} b_{i\bj})' \right) \\
\bp \given  \cdot &\sim Dirichlet(\sum_{\bj} b_{1\bj} + \delta, \ldots, \sum_{\bj} b_{C\bj} + \delta)
\end{align*}
Here $\odot$ denotes the Hadamard (elementwise) product.

Finally, as in Section \ref{sec:bayes}, we update $\gamma\kpow_i$'s using a metropolis random walk with log-normal proposal to sample from the full conditionals 
\begin{align*}
p(\gamma\kpow_i \given \cdot) \propto & \frac{\Gamma(C\gamma\kpow_i\epsilon + \gamma_i)}{\Gamma(\gamma\kpow_i\epsilon)^{C - 1}\Gamma(\gamma\kpow_i\epsilon + \gamma\kpow_i)} \times \\
& (\gamma\kpow_i) ^{\alpha-1} \exp(-\beta \gamma\kpow_i) \prod_{j}(m\kpow_{ij})^{\gamma\kpow_i\epsilon +  \gamma\kpow_i1(i  = j) } .
\end{align*}

\subsection{Individual level classifications}\label{sec:enscod} As illustrated in Section \ref{sec:cod}, the ensemble transfer learner can also predict the individual-level class memberships. Using Bayes theorem we have
\begin{align*}
p(G(\bs_r)=i \given a_r^{(1)}=j_1, \ldots, a_r\Kpow = j_k) = \frac 1{\sum_\bj u_{i\bj} p_i } u_{i\bj} p_i .
\end{align*}
Since posterior distributions of $u_{i\bj}$'s and $\bp$ have already been sampled, we can generate posterior samples of $G(\bs_r)$ post-hoc using the composition sampling approach demonstrated in Section \ref{sec:cod}.

For the independent ensemble model, one can  recover the posterior distribution of the individual class memberships in the exact same way. Only additional step would be to first calculate the $\bu_\bj$'s as they are no longer part of the Gibbs sampler. 

\section{Detailed analysis of the simulation results}\label{sec:simdetails} In this Section we present a much more thorough analysis of the simulation study, as well as investigate additional methods to generate population-level class probabilities in the target domain. We first present the exact choices of $\bM$ used. We have $\bM_1=\bI$, 
\begin{equation*}
\bM_2 = \left( \begin{array}{cccc}
1.00 & 0 &  0 &  0\\
0.65 & 0.35 & 0 & 0\\
0 & 0 & 0.5 & 0.5\\
0 & 0 & 0 & 1
\end{array} \right)
\end{equation*}
and $\bM_3 = 0.6 * \bI+ 0.1* \bone \bone'$. The first choice represents the case where the algorithm $A$ is perfect for predicting in the target domain. The specific choice of $\bM_2$ depicts the scenario that $65\%$ of Diarrhea/Dysentry cases are classified as Pneumonia and $50\%$ of Sepsis deaths are categorized as some other cause. Finally, $\bM_3$ represents the scenario where there are many small misclassifications. 

\subsection{Impact of difference in marginal distributions between source and target domains}\label{sec:simcsmf}   
\begin{figure}
	\centering
	\includegraphics[scale=0.75]{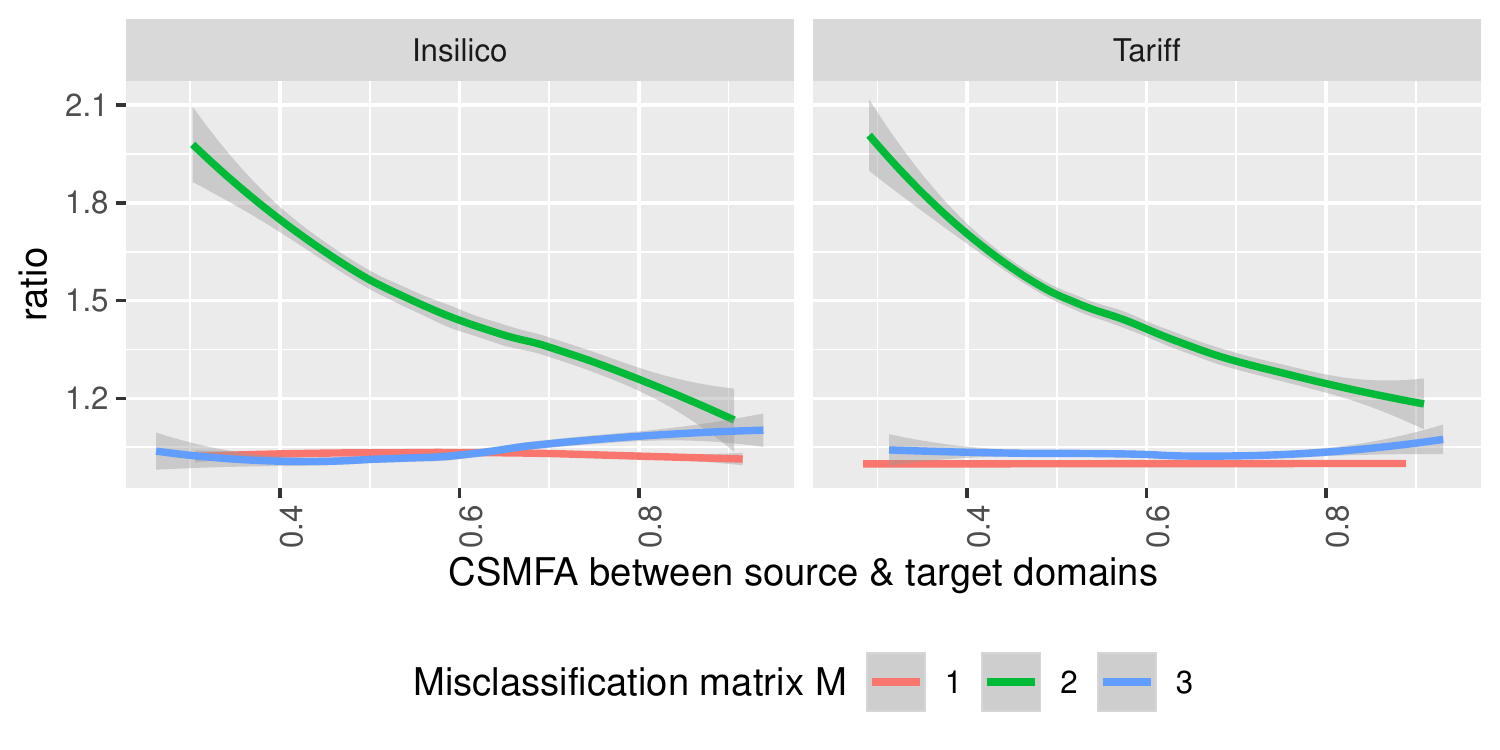}
	\vskip -7mm \caption{Ratio of CSMFA of domain-adapted and baseline CCVA}\label{fig:rev}
\end{figure}

We first investigate how the performance of our Bayesian transfer learning model is impacted by the disparity in class distribution between $\calD_S$ and $\calD_T$. Figure \ref{fig:rev} plots the smoothed ratio of the CSMFA of the baseline estimates and their calibrated analogs from our model, as a function of the true CSMFA between the class probabilities $\bp_\calG$ and $\bp_\calU$ in the source and target domains. The left panels correspond to data generated using InsilicoVA and hence assesses the performance of Insilico$_G$ and $\ic$ by plotting the ratio CSMFA$(\ic)/$ CSMFA$(\ig)$. Similarly, the right panels correspond to data generated using Tariff and compares the estimates from $\tg$ and $\tc$. We only present the results for $n=400$, as we will discuss the role of $n$ in Section \ref{sec:simsize}. %The four rows corresponds to four sizes of the hospital set $\calL$ and and three colors correspond to data generated using the three different misclassification matrices.
We first note that when data was generated using the misclassification matrix $\bM_1=\bI$, the ratio is exactly one. This corroborates the result in Theorem \ref{th:postp} that if $A$ classifies instances flawlessly in $\calD_T$, then the baseline and domain adapted estimates are same. For $\bM_3$, i.e. when the misclassification rate is small, the ratio is also close to one with the domain-adapted estimate being slightly more accurate in general. For $\bM_2$, which portrays the scenario where the the baseline learner trained on source domain is systematically and substantially biased, one can clearly see the benefit of transfer learning. The CSMFA is significantly better after domain-adaptation. It also nicely shows the utility of transfer learning as a function of $x = CSMFA(\bp_\calG,\bp_\calU)$. Unsuprisingly, the ratio is decreasing with increasing $x$. When $x$ is small, i.e., there exists much disparity in the marginal class distributions between the source and target domains, the ratio is close to two, implying that domain-adaptation yields near $100\%$ gain in accuracy. When $x$ is close to one, the improvement is much less stark, which is expected as in this scenario the class probabilities in the non-local and local populations are almost identical.

\subsection{Biases in estimates of probabilities for each class}\label{sec:bias}
\begin{figure}[h]
	\centering
	\includegraphics[scale=0.75]{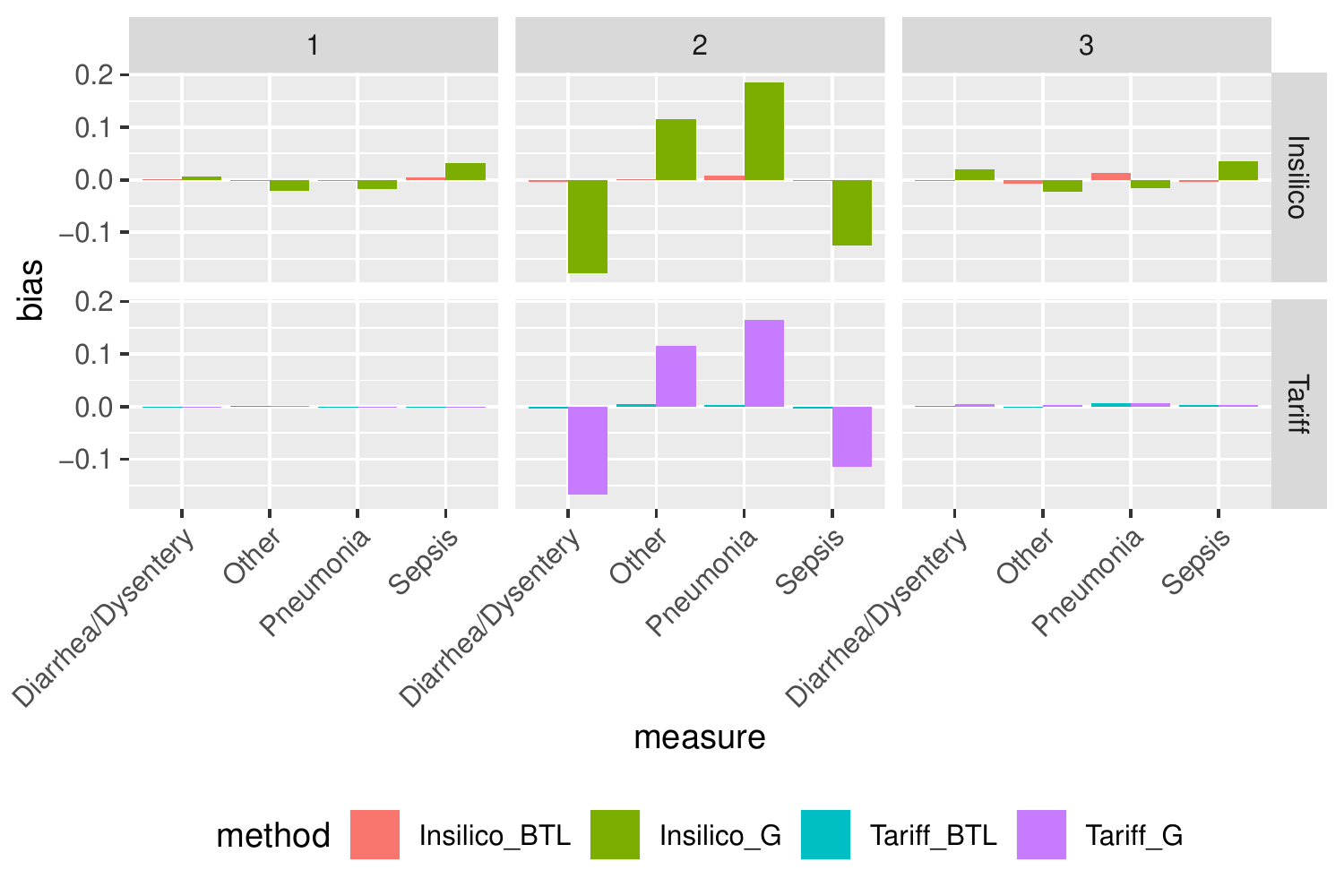}
	\vskip -7mm \caption{Biases in the average estimates of individual cause prevalences}\label{fig:bias}
\end{figure}

We also look at the biases in the estimates of each of the four class probabilities in Figure \ref{fig:bias}. The top and bottom rows correspond to data generated using InsilicoVA and Tariff respectively. The three columns correspond to three choices of $\bM$. We see that there is almost no bias for $\bM_1$ for all the methods, for $\bM_3$ the baseline $\tg$ estimates are generally unbiased, whereas the baseline $\ig$
show small biases. However, for $\bM_2$ we see the substantial biases in the estimates from both the baseline approaches. As expected due to the specification of $\bM_2$, the baseline learners underestimate $P( Diarrhea/Dysentry)$ and $P(Sepsis)$ while overestimating $P(Pneumonia)$ and $P(Other)$ are overestimated. The domain-adapted estimates $\tc$ and $\ic$ are unbiased for all the settings. 

\subsection{Role of limited labeled data in target domain}\label{sec:simsize} We now investigate the role of the sample size $n$ and the marginal class distribution $\bp_\calL$ of $\calL$. Additionally, as an alternate to our transfer learning approach, we also consider including the local labeled data $\calL$ as part the training data for the CCVA algorithms. So, we have four more methods $\tch$, $\tgh$, $\ih$ and $\igh$, where the sub-scripts indicate the training data used. 

\begin{figure}
	\centering
	\includegraphics[scale=0.75]{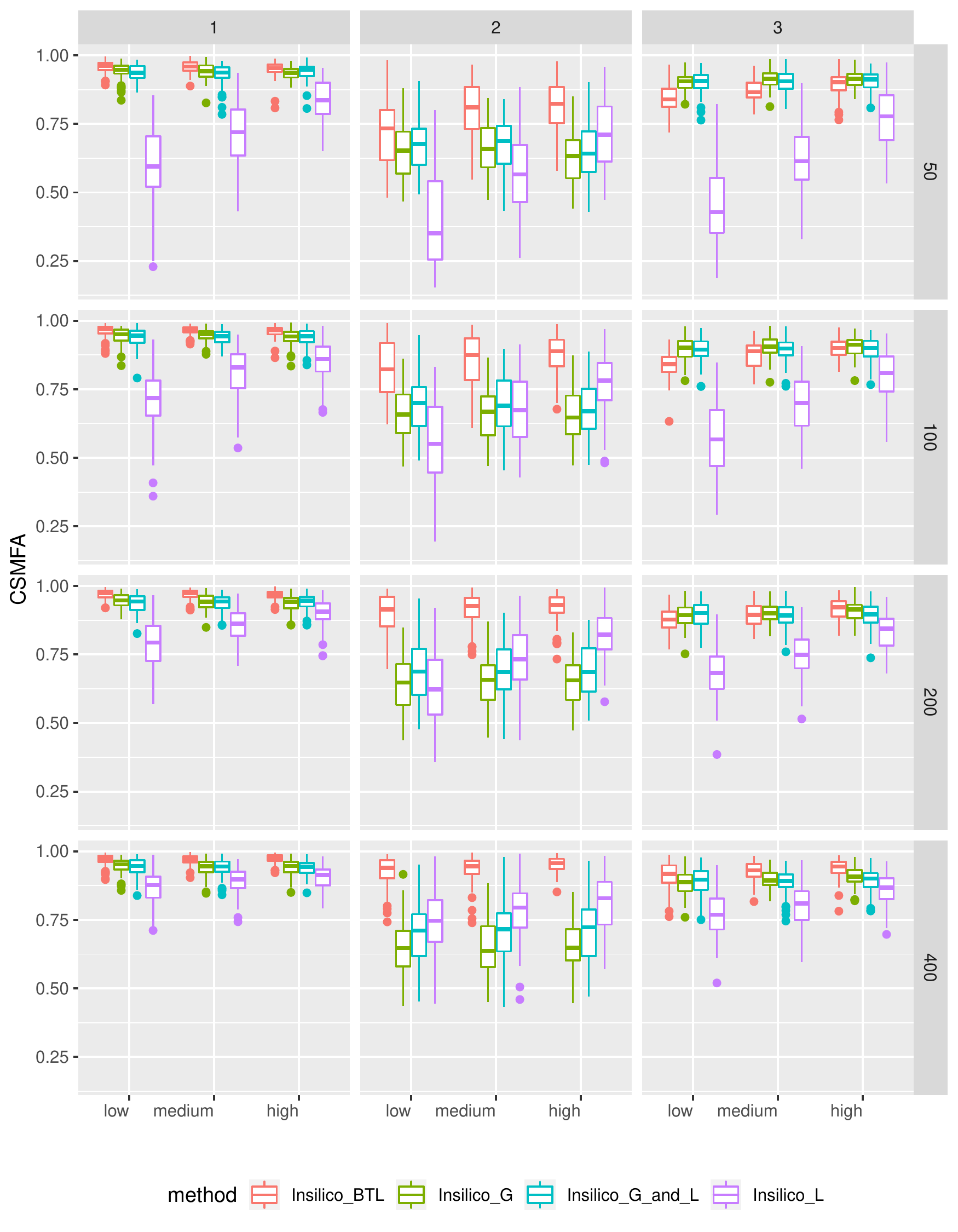}
	\vskip -7mm \caption{CSMFA for additional methods when data is generated using InsilicoVA}\label{fig:csmfins}
\end{figure}

When data is generated using InsilicoVA, Figure \ref{fig:csmfins} provides the boxplots of CSMF accuracy of the methods for all the scenarios as a function of $n$ (rows), choice of $\bM$ (columns) and $\rho$ --- the CSMFA-range between $\bp_\calL$ and $\bp_\calU$ ($x$-axis in each sub-figure). The analogous results for data generated using Tariff, provided in Figure \ref{fig:csmftar} of the supplement, reveals exactly similar trends. 

We unpack many different conclusions from this Figure. First we look at the performances of $\ig$ and $\ic$. These two methods was already compared in Section \ref{sec:simcsmf}, but only for fixed $n=400$ and averaged across all $\rho$. Here, further analyzing the performances as a function of $n$ and $\rho$, we see that the CSMFA of calibrated VA using our model increases with increase in $n$. Also, there is a drastic gain in precision of the calibrated estimates with the confidence bands shortening with increase in $n$ from $50$ to $400$. Additionally, we see that the CSMFA for $\ic$ increases as $\rho$ goes from $low$ to $medium$ to $high$, although the gain is not as drastic. This indicates that the transfer learning procedure, is reasonably robust to the value of $\rho$, but does benefit to a small extent from improved concordance between the class probabilities in  $\calL$ and $\calU$. 
Of course, $\ig$ is not affected by either $n$ or $\rho$. In general, for $\bM_3$, we see that only when both $n$ is small and $\rho$ is $low$, the $\ig$ produces slightly better estimates than $\ic$. For all other cases, $\ic$ yields higher or similar CSMF. For $\bM_2$, we see $\ic$ dominates $\ig$ across all scenarios. The gains from increase in $n$ and $\rho$ are evident here as well. Finally, for $\bM_1$, the performance of $\ic$ is identical to $\ig$, as is guaranteed by Theorem \ref{th:postp}, and is not affected by $n$ or $\rho$. 

Next, we look at the performance of $\ih$ and $\igh$. For $\bM_1$ and $\bM_3$, $\ih$ performs quite poorly, generally producing the lowest CSMF. $\ih$ is also highly sensitive to both $\rho$ and $n$, yielding highly variable and inaccurate estimates for $low$ $\rho$ and $n$, and improving sharply as either increases. Only for $\bM_2$, for large $n$ or large $\rho$, it does better than $\ig$. As this setting\ portrays substantial difference in the conditional distributions between the source and target population, $\ih$, trained on local data, does better. CSMFA from $\igh$, which uses both the source and target labeled data in the training, generally lies between the CSMFA from $\ig$ and $\ih$, and is much closer to the former as $\calG$ far outnumbers $\calL$. Finally, comparing $\ih$ and $\igh$ to $\ic$, we see that the latter does substantially better uniformly across the scenarios. This shows that with a small labeled dataset in $\calD_T$, our transfer learning approach is a more resourceful way of exploiting this limited data and results in more accurate and robust estimates. 

\subsection{Comparison with the naive transfer learning}\label{sec:simnaive}
\begin{figure}[t]
	\centering
	\includegraphics[scale=0.75]{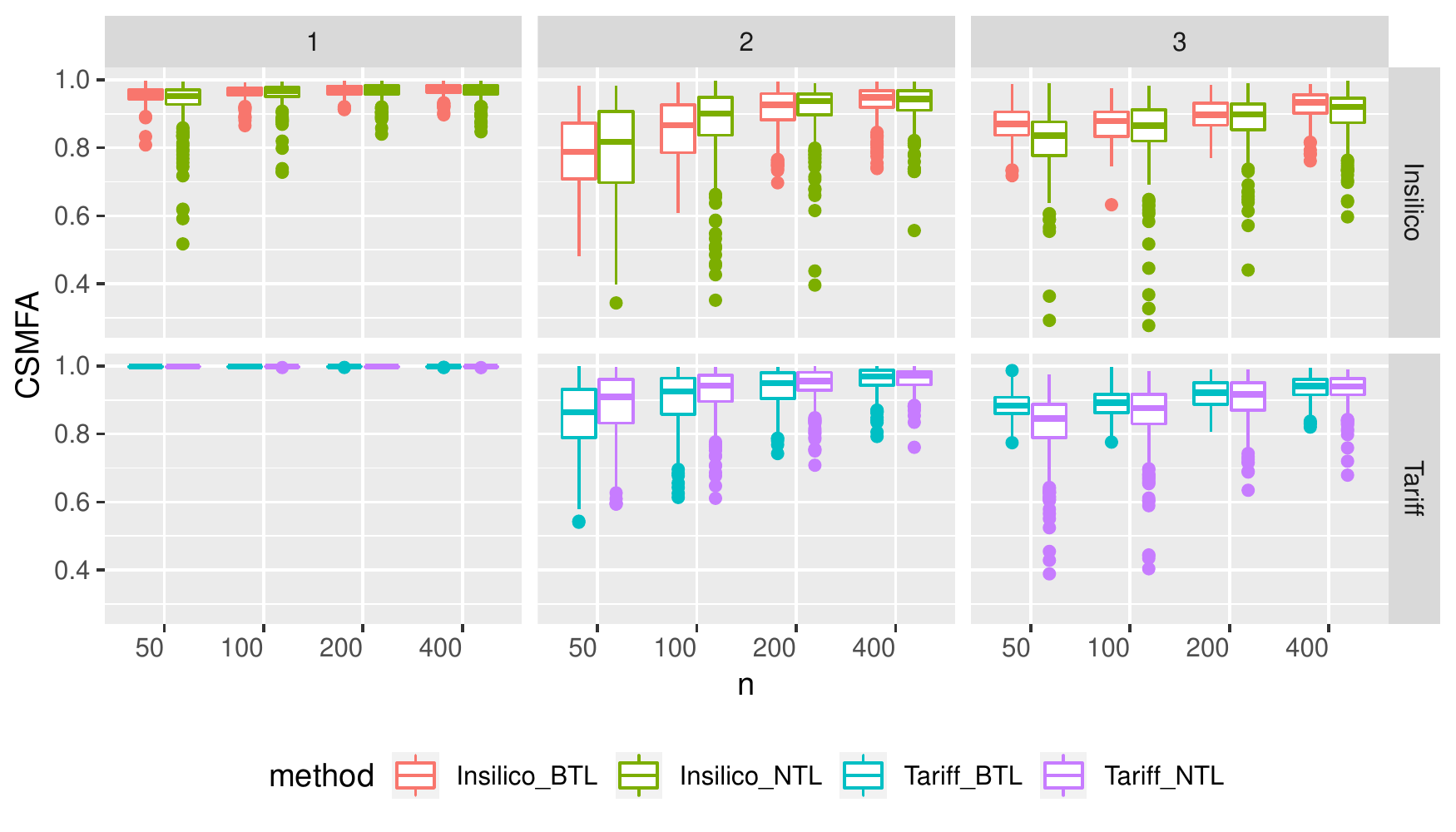}
	\vskip -7mm \caption{CSMFA of naive and Bayesian transfer learning}\label{fig:naive}
\end{figure}

To understand the importance of the Bayesian regularization used in the the transfer learning, we also compare with the naive transfer learning based on MLE, outlined in Section \ref{sec:cal}. 
We refer to the naive transfer learning using Tariff and InsilicoVA respectively as $\tnc$ and $\inc$. Figure \ref{fig:naive} compares the CMSFA for the naive and Bayesian regularized transfer learning approaches. 
Once again, the top and bottom row corresponds to data generated using Tariff and InsilicoVA respectively, the three columns are for three choices of $\bM$ and within each setting, we plot the boxplots of CSMFA as a function of $n$.  

We see that, generally the median estimates from the naive approach is similar to the ones produced using the Bayesian regularized analog. However, there is notable difference in the variability of CSMFA, with the naive approach producing a wide range of values with several extreme estimates. The problem is exacerbated for smaller values of $n$. The results from the Bayesian model are more stable with uniformly lesser variation across all the settings. It is evident, that in real data analysis, where the truth is unknown, the Bayesian model will be much more reliable than the MLE based solution which seems to be quite likely to yield absurd estimates. 

\subsection{Performance of ensemble models}\label{sec:simens} We now analyze the performance of the joint ($\ej$) and independent ($\ei$) ensemble transfer learning models introduced in Section \ref{sec:ens}. These models use output from both Tariff and InsilicoVA whereas the single-classifier models $\tc$ and $\ic$ only use the output from one CCVA algorithm. For a given dataset, we define 
\begin{align*}
\delta = & \max(\mbox{CSMFA}(\ic),\mbox{CSMFA}(\tc)) - \\
& \min(\mbox{CSMFA}(\ic),\mbox{CSMFA}(\tc)) .
\end{align*} 
In other words, $\delta$ denotes the difference in CSMFA of the calibrated VA using the most and least accurate classifiers. A small $\delta$ implies transfer learning with either of the baseline classifiers yield similar results, whereas larger values of $\delta$ clearly insinuate that transfer learning with one of the baseline classifiers is more accurate than the other one. For an ensemble method that aims to guard against inclusion of an inaccurate method, one would expect that CSMFA for the ensemble method should be closer to that of the best performing method. Equivalently, if 
\[ \nu = CSMFA(Ensemble) - \min(\mbox{CSMFA}(\ic),\mbox{CSMFA}(\tc)), \] 
where Ensemble refers to either $\ei$ or $\ej$, 
then $\nu$ should be greater than $\delta /2$. 

\begin{figure}[t]
	\centering
	\includegraphics[scale=0.6]{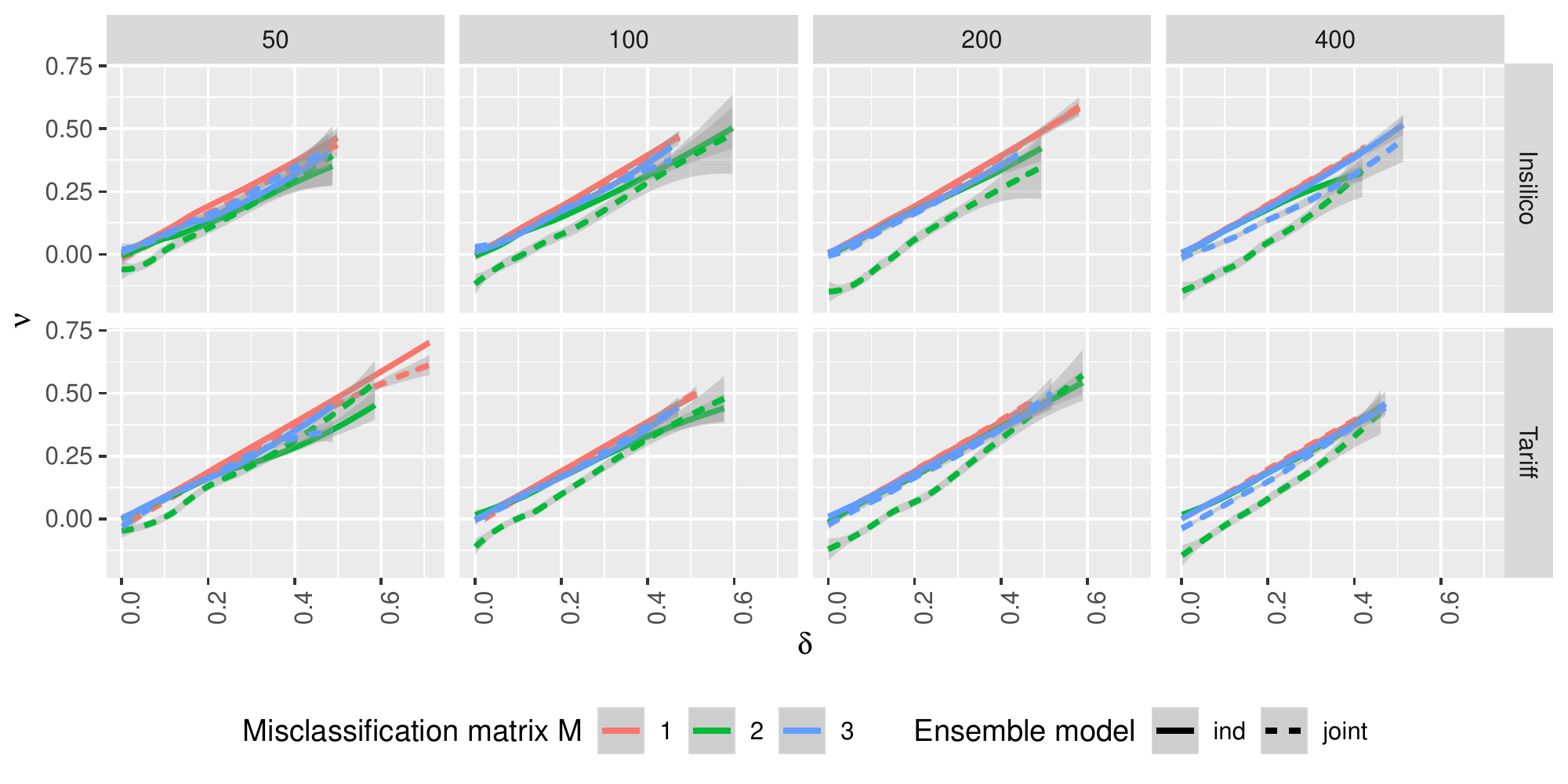}
	\caption{Performance of the ensemble models}\label{fig:ens}
\end{figure}

Figure \ref{fig:ens} plots $\nu$ as a smoothed function of $\delta$. We first note that, for $\bM_1$ (red lines), the $(\nu,\delta)$ curve for the joint sampler nearly coincides with the $45$-degree line. Since, in our data generation process, under $\bM_1$ one of the classifiers is fully accurate, this empirically verifies the theoretical guarantee in Theorem \ref{th:joint}, that in such settings posterior mean of class probabilities from the ensemble approach is same as that from the best classifier. While the independent ensemble model does not enjoy this theoretical property, in practice we see that for $\bM_1$, $\nu$ is also identical  to $\delta$. 
For $\bM_2$ and $\bM_3$, across all scenarios, $\nu$ is close to $\delta$, i.e. estimates from both the $\ej$ and $\ei$ models generally aligns much closer to the best performing single-classifer transfer learner. There are no significant trends with respect to either the size of $\calL$ ($n$) or the data generating algorithm -- Insilico (top-row) and Tariff (bottom-row).
The $\ei$ model seems to do slightly better than the joint model. Since, it is also the faster model, we only use this version of the ensemble model for subsequent analysis. The performance of the ensemble samplers is quite reassuring especially for larger $\delta$, as it demonstrates the robustness to inclusion of a bad method via averaging over multiple methods. As $\nu$ seems to be substantially greater than $\delta/2$ for most of the curves, it also shows why our model based method averaging is superior to simply taking average of the estimated class-probabilities from the different methods, which is much more affected by the worst method. 

\subsection{Individual level classification}\label{sec:simcod}
\begin{figure}
	\centering
	\includegraphics[scale=0.75]{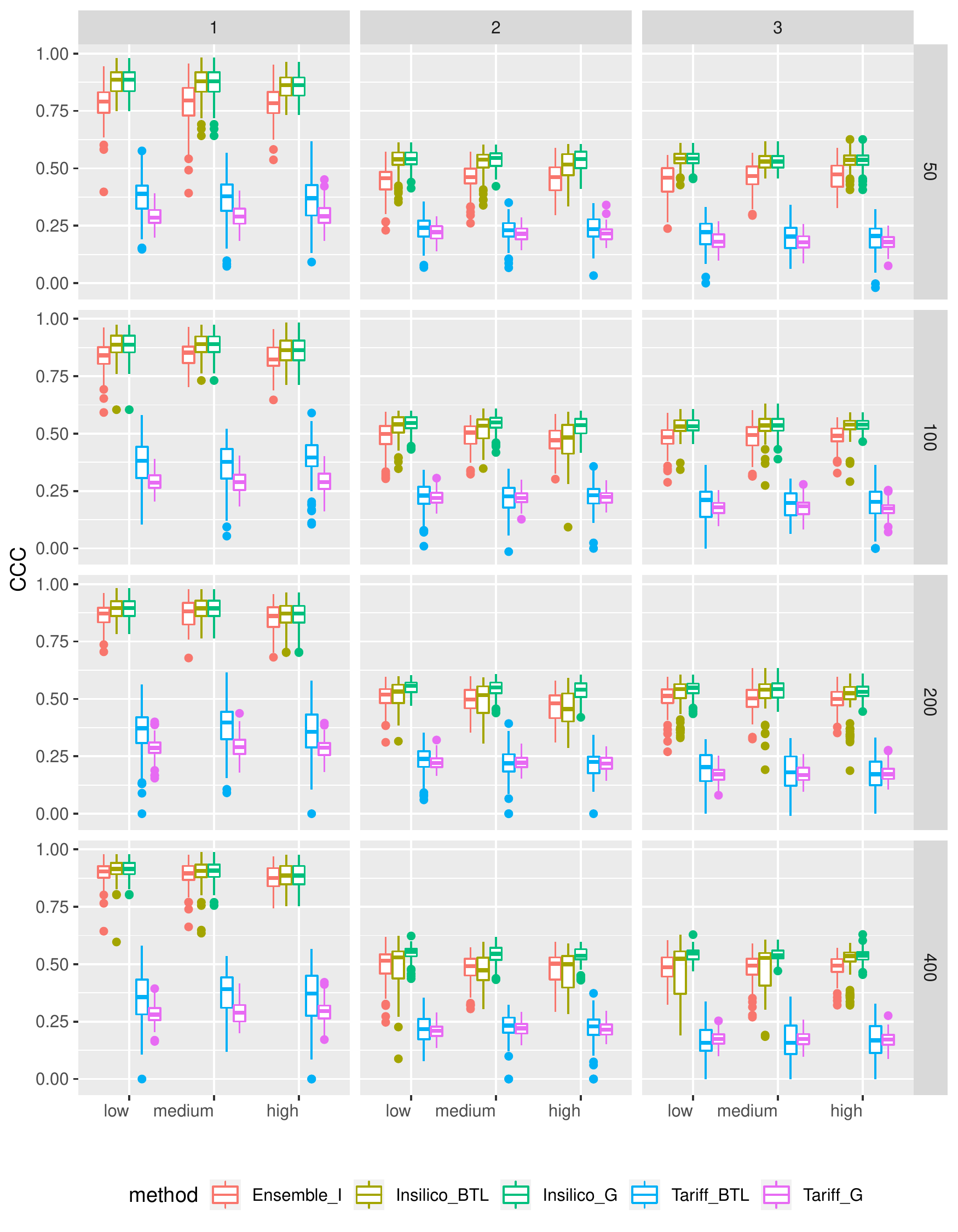}
	\vskip -7mm \caption{CCC when data is generated using InsilicoVA}\label{fig:cccins}
\end{figure}

\begin{figure}
	\centering
	\includegraphics[scale=0.75]{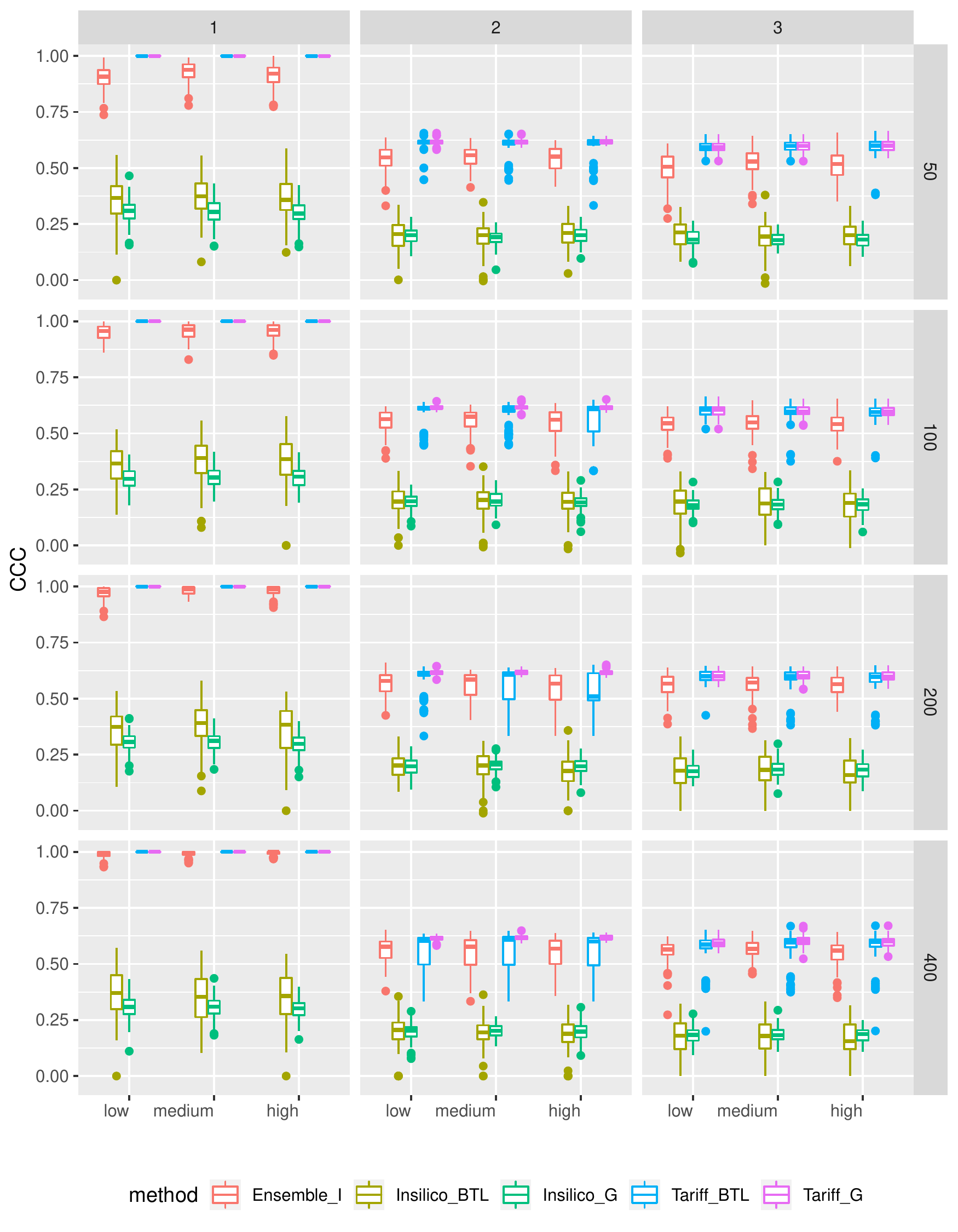}
	\vskip -7mm \caption{CCC when data is generated using Tariff}\label{fig:ccctar}
\end{figure}

As mentioned earlier, predicting individual classes is not our primary goal. Nonetheless, we have outlined a simple way to obtain individual predictions using our transfer learning model. Here we compared its accuracy using the Chance Corrected Concordance \citep{murray2011robust} defined as
\[ \mbox{CCC} = \frac 1C \sum_{i=1}^C \frac{\frac{TP_i}{TP_i+TN_i} - \frac 1N}{1- \frac 1N}\]
where $TP_i$ and $TN_i$ denote the true positive and true negative rates for class $i$. We only analyze the case when when the data is generated using Insilico (Figure \ref{fig:cccins}). The roles are simply reversed when data is generated using Tariff (Figure \ref{fig:ccctar}). 
We see in Figure \ref{fig:cccins} that CCC for $\ig$ and $\ic$ are better than those of $\tg$ and $\tc$ respectively. This is expected as analyzing data using the true model is expected to perform better than the misspecified model. CCC from the domain-adapted ($\ic$) and baseline ($\ig$) versions of the same CCVA algorithm, which was used in data generation, were similar. For the misspecified model, the domain-adapted  $\tc$ produced slightly better CCC than its baseline analog $\tg$. However, these gains in CCC from our transfer learning model are not substantial and perhaps indicate that if individual prediction is of interest then more advanced methods need to be considered than the simple approach we have outlined. However, even using our crude approach, we see that the ensemble model ($\ei$) produces CCC closer to $\ic$ and $\ig$, and much better than the CCC obtained by both $\tc$ and $\tg$. This once again furnishes evidence of the robust performance of the ensemble model, and in practice, when we will not know which algorithm works best, using the ensemble model will safeguard against choosing a bad algorithm.

\section{Sensitivity analysis for India and Tanzania}\label{sec:sens}

We employed a resampling approach similar to \citep{murray2011population} in order to perform a sensitivity analysis of whether the (empirically determined) specific CSMFs $\bp_\calG$ and $\bp_\calU$ respectively of the source and target domains in the PHMRC data impacted the data analysis in Section \ref{sec:phmrc}. To sample data into $\calG$ and $\calU$, we drew $\bp_\calG$ and $\bp_\calU$ from independent uninformative Dirichlet distributions. We sampled 800 subjects from outside of the country of interest, with replacement according to probability $\bp_\calG$ and based on their GS-COD, into $\calG$. A similar process was used to create $\calU$, except using records from exclusively within the country of interest and using the probability vector $\bp_{\calU}$. For creating $\calL$, we sampled $n$ subjects from within the country of interest with replacement, using the empirical distribution of COD within that country. 
We used $\calG$ to train models $Tariff_{G}$, $Insilico_{G}$ and calibrated them using $\calL$ to get $\tc$, $\ic$, and $Ensemble_{I}$. We used the  $\bp_{\calU}$ drawn from the Dirichlet distribution as the true CSMF and computed the CSMFA for all the models. This whole process was repeated 500 times for each country and value of $n$. 

\begin{figure}[h!]
	\centering
	\includegraphics[scale=0.2]{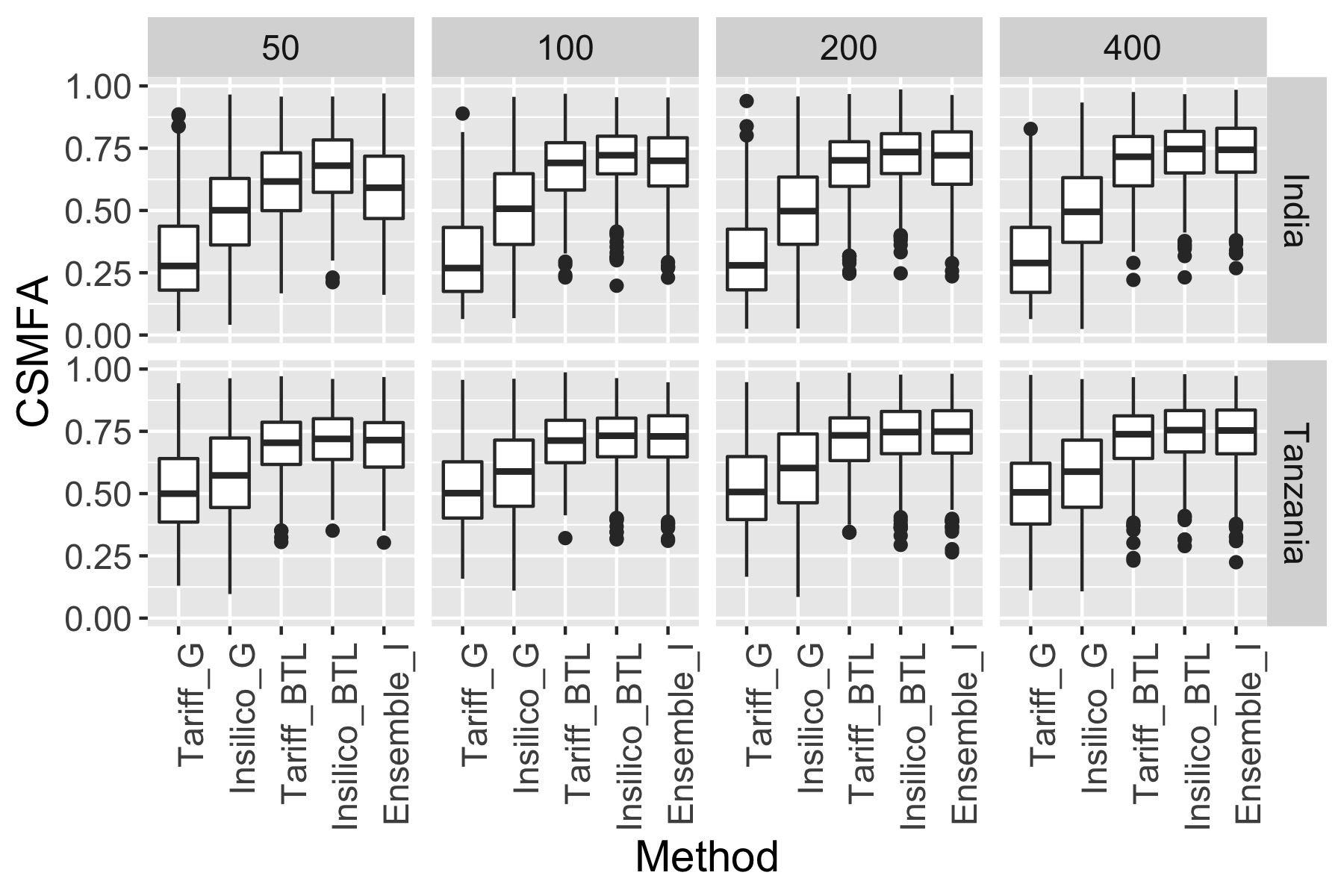}
	\vskip -7mm \caption{Sensitivity analysis of CSMFA of competing algorithms for the top 3 COD + others}\label{fig:sensanalysis}
\end{figure}

We again can visualize the results in Figure \ref{fig:sensanalysis}. First observation is that the CSMFA of all algorithms is dramatically lower than in Section \ref{sec:phmrc} and have higher uncertainty. This is expected as we are now averaging over $500$ pairs of difference cause distributions for the training and test set. We see that both in India and Tanzania the domain-adapted methods $\tc$ and $\ic$ perform much better than the baseline analogs $\tg$ and $\ig$ respectively, with the gain in CSMFA becoming more substantial with increase in $n$.  
For India, it appears that except for $n=50$, $Ensemble_{I}$ generally performs similar to $\ic$ and  better than $\tc$. For Tanzania, all three transfer learning models perform similarly, and are better than the baseline models. 
Overall, the sensitivity analysis shows that over a large range of possible CSMF compositions for $\calG$ and $\calU$, the calibrated estimates of $\bp_\calU$ were more accurate.

\subsection{Additional figures}\label{sec:supfig}
\begin{figure}
	\centering
	\includegraphics[scale=0.75]{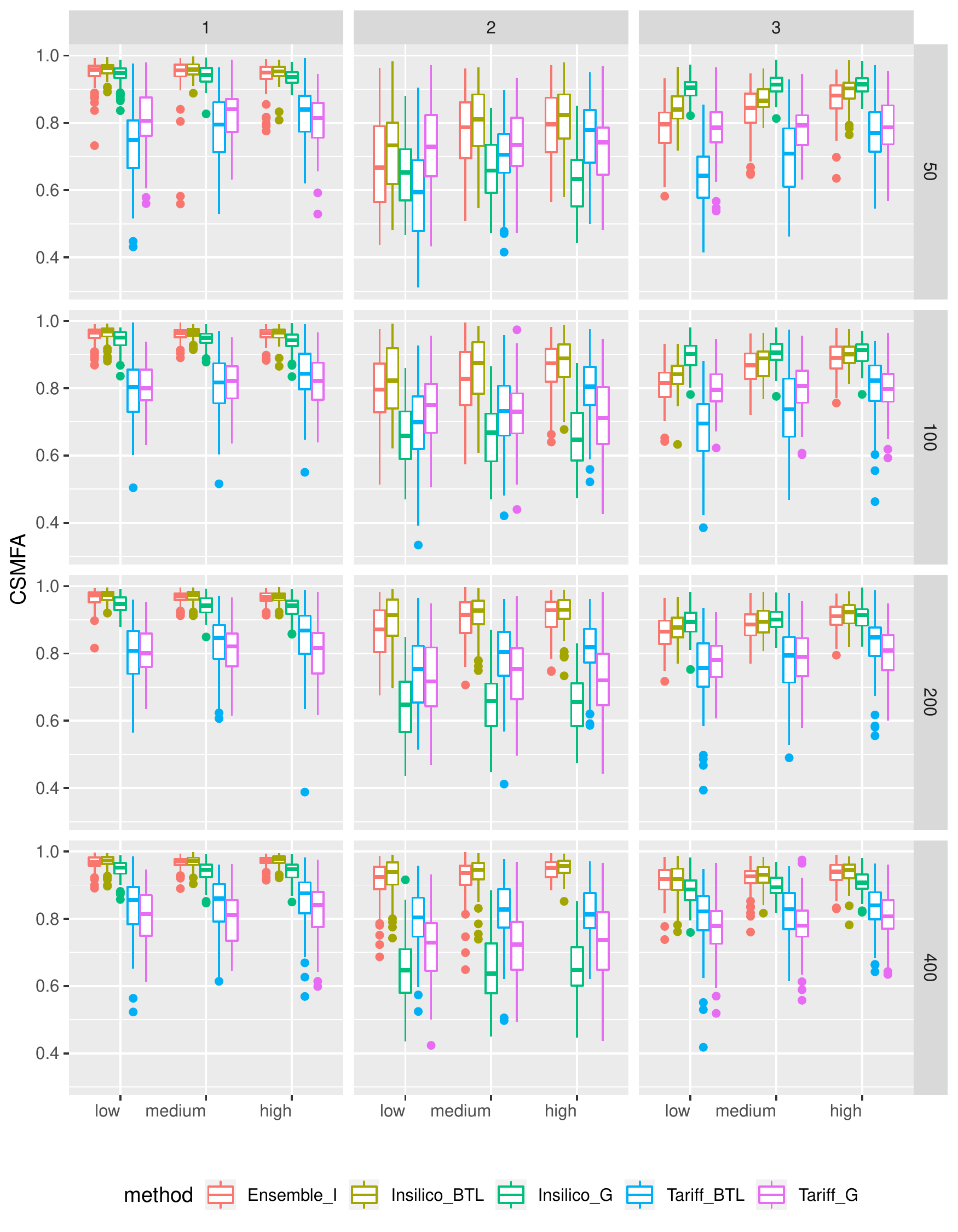}
	\vskip -7mm \caption{CSMF when data is generated using InsilicoVA}\label{fig:csmfensins}
\end{figure}

\begin{figure}
	\centering
	\includegraphics[scale=0.75]{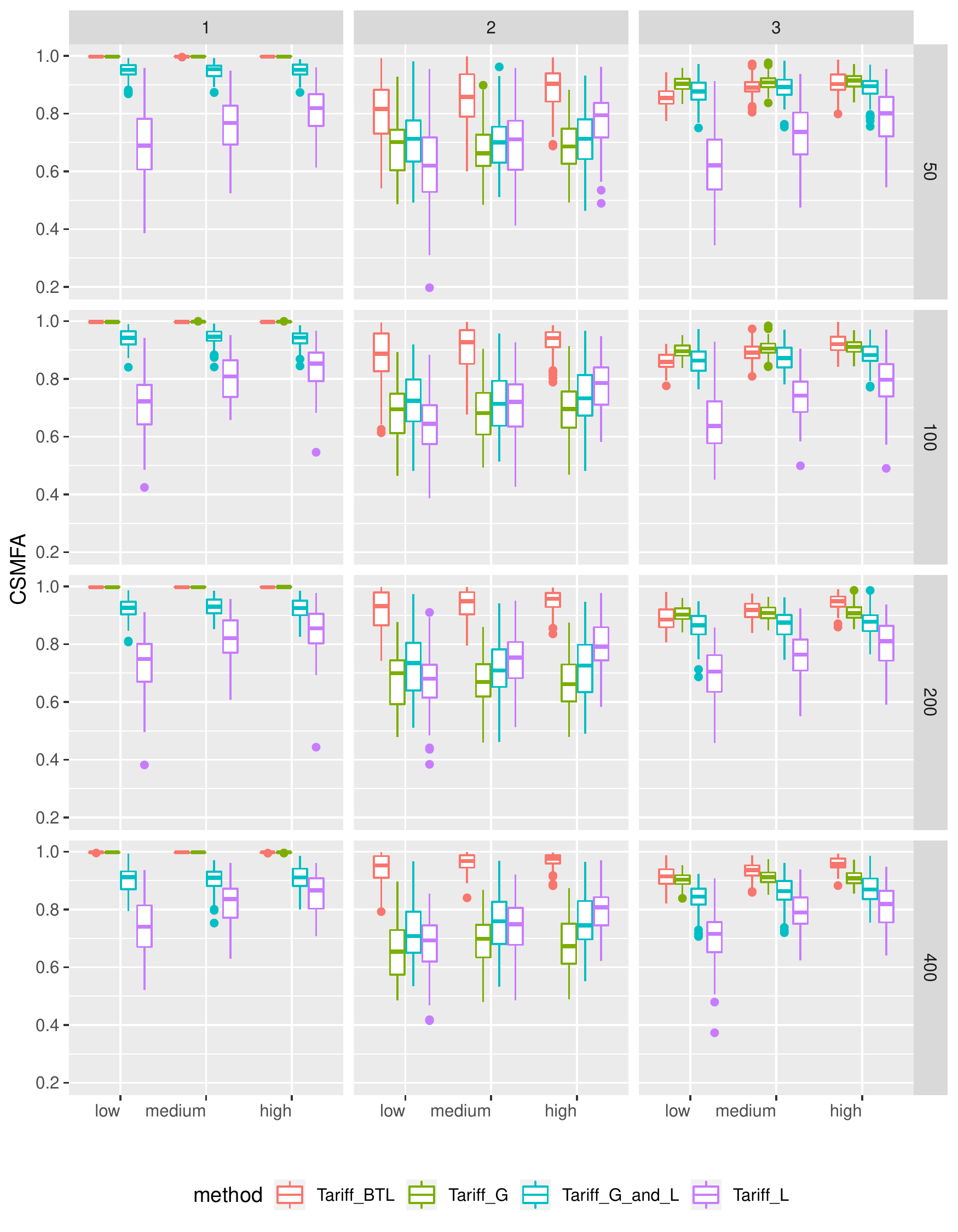}
	\vskip -7mm \caption{CSMF for the additional methods of Section \ref{sec:simsize} when data is generated using Tariff}\label{fig:csmftar}
\end{figure}

\bibliographystyle{chicago}
\bibliography{references}

\end{document}